\documentclass[referee]{aa} 
\usepackage{graphicx}
\usepackage{subeqnarrayart}
\begin{document}
%
%
%
\newenvironment{lefteqnarray}{\arraycolsep=0pt\begin{eqnarray}}
{\end{eqnarray}\protect\aftergroup\ignorespaces}
\newenvironment{lefteqnarray*}{\arraycolsep=0pt\begin{eqnarray*}}
{\end{eqnarray*}\protect\aftergroup\ignorespaces}
\newenvironment{leftsubeqnarray}{\arraycolsep=0pt\begin{subeqnarray}}
{\end{subeqnarray}\protect\aftergroup\ignorespaces}
\newcommand{\diff}{{\rm\,d}}
\newcommand{\appleq}{\stackrel{<}{\sim}}
\newcommand{\appgeq}{\stackrel{>}{\sim}}
\newcommand{\Int}{\mathop{\rm Int}\nolimits}
\newcommand{\Nint}{\mathop{\rm Nint}\nolimits}
\newcommand{\arctg}{\mathop{\rm arctg}\nolimits}

\title{The potential-energy tensors for subsystems. IV.
}

\subtitle{Homeoidally striated density profiles with 
a central cusp.
}
\authorrunning{R. Caimmi and C. Marmo}
\titlerunning{density profiles with a central cusp}

\author{R. Caimmi
\and C. Marmo}

\offprints{R. Caimmi}

\institute{Dipartimento di Astronomia, Universit\`a di Padova,   \\
	      Vicolo Osservatorio 2, I-35122 Padova, Italy\\
caimmi@pd.astro.it -- marmo@pd.astro.it
	      }

\date{Received: ............................$\qquad$  Accepted: .................................}

\abstract{A general theory of homeoidally striated density
profiles where no divergence occurs, is adapted to cuspy
density profiles, with a suitable choice of the scaling
density and the scaling radius.   A general formulation of some physical
parameters, such as angular-momentum vector, rotational-energy
tensor (both calculated in connection with a special class of rotational
velocity fields), inertia tensor, and self potential-energy
tensor, is performed.   Other potential-energy tensors involving
two density profiles where the boundaries are similar and
similarly placed, are also expressed.    Explicit results are
attained for three special cases of physical interest: NFW (e.g.,
Navarro et al. 1997) and MOA (e.g., Moore et al. 1999)
density profiles, which fit to a good extent the
results of high-resolution simulations for dark matter haloes, 
and H (Hernquist 1990) density profiles,
which closely approximate the de Vaucouleurs $r^{1/4}$ law
for elliptical galaxies.   The virial theorem in tensor form
for two-component systems
is written for each subsystem, and applied to
giant elliptical galaxies.   The predicted
velocity dispersion along the line of sight, in 
the limiting case where a principal axis points 
towards the observer, is found to be
consistent with observations except for (intrinsic) $E7$ 
configurations where the major axis points towards the observer.   
If dark matter haloes host an amount of undetected baryons about twice as 
massive as the stellar subsystem, and undetected 
baryons trace non baryonic matter therein, two main consequences 
arise, namely (i) velocity dispersions along the line of sight are 
lower than in absence of undetected baryons, and (ii) dark matter 
haloes are dynamically ``hotter'' than stellar
ellipsoids, the transition occurring when the amount of
undetected baryons is about one and a half times that of the stellar subsystem.
In this view, both the observation that the temperature of the
extended hot gas exceeds the central stellar temperature, and
the fact that the non baryonic matter is dynamically ``hotter''
than the stars, are a reflection of the presence
of undetected baryons, which trace the dark halo and are about twice as 
massive as the stellar ellipsoid.
   }

   \maketitle
%

\section{Introduction}\label{intro}
According to standard CDM or $\Lambda$CDM cosmological scenarios, 
large-scale celestial objects such as galaxies and clusters of 
galaxies, are made of at least two components: one, baryonic and
more concentrated, embedded within one other, non baryonic and
dissipationless, usually named dark matter halo.
After a wide number of both analytical and numerical
studies (e.g., Cole \& Lacey 1996; Navarro et al. 1995, 1996, 
1997; Moore et al. 1998, 1999; Fukushige \& Makino 2001; Klypin 
et al. 2001), it has been realized that dark
matter haloes which virialize from hierarchical clustering 
show universal density profiles, $\rho=\rho(r; \rho^\dagger, r^\dagger)$,
where $\rho^\dagger$ is a scaling density and $r^\dagger$ is a 
scaling radius.    In this view, smaller haloes formed first from
initial density perturbations and then merged with each 
other, or were tidally disrupted from previously formed 
mergers, to become larger haloes.

The density profile is (i) self-similar, in the sense 
that it has the same expression, independent of time
(e.g., Fukushige \& Makino 2001), and (ii) universal, in the 
sense that it
has the same expression, independent of halo mass,
initial density perturbation spectrum, or value of
cosmological parameters (e.g., Navarro et al. 1997; Fukushige
\& Makino 2001).   A
satisfactory fit to the results of numerical simulations
is the family of density profiles (e.g., Hernquist 1990; 
Zhao 1996):
\begin{equation}
\label{eq:runi}
\rho\left(\frac r{r^\dagger}\right)=\frac{\rho^\dagger}{(r/r^\dagger)^\gamma
[1+(r/r^\dagger)^\alpha]^\chi}~~;\quad\chi=\frac{\beta-\gamma}
\alpha~~;
\end{equation}
for a suitable choice of exponents, $\alpha$, $\beta$, 
and $\gamma$.

This family includes both cuspy profiles first proposed by
Navarro et al. (1995, 1996, 1997), $(\alpha,\beta,\gamma)=
(1,3,1)$, and the so called modified isothermal profile,
$(\alpha,\beta,\gamma)=(2,2,0)$, which is the most widely
used model for the halo density distribution in analyses
of observed rotation curves.   It also includes the perfect
ellipsoid (e.g., de Zeeuw 1985), $(\alpha,\beta,\gamma)=
(2,4,0)$, which is the sole (known) ellipsoidal density
profile where a test particle admits three global integrals
of motion.   Finally, it includes the Hernquist (1990)
density profile, $(\alpha,\beta,\gamma)=(1,4,1)$, which
closely approximates the de Vaucouleurs $r^{1/4}$ law
for elliptical galaxies.   In 
dealing with the formation of dark matter haloes from 
hierarchical clustering in both CDM and $\Lambda$CDM scenarios,
recent high-resolution simulations allow $(\alpha,\beta,\gamma)=
(3/2,3,3/2)$, as a best fit (e.g., Ghigna et al. 2000; 
Fukushige \& Makino 2001; Klypin et al. 2001), as first 
advocated by Moore et al. (1998, 1999)%
\footnote{More precisely, an exponent $\alpha=1.4$ was derived
by Moore et al. (1998), while the value $\alpha=1.5$ was
established by Moore et al. (1999).}.
%

Though Eq.\,(\ref{eq:runi}) implies null density at infinite
radius, the mass distribution has necessarily to be truncated
for two types of reasons.   First, the presence of
neighbouring systems makes the tidal radius an upper limit.
On the other hand, isolated, over-dense objects cannot extend 
outside the Hubble sphere of equal mass.
Second, the total mass, deduced from Eq.\,(\ref{eq:runi}) for
an infinitely extended configuration, is divergent, at least
with regard to the special choices of exponents, $(\alpha,
\beta,\gamma)=(1,3,1)$, hereafter quoted as NFW density
profile, and $(\alpha,\beta,\gamma)=(3/2,3,3/2)$, hereafter
quoted as MOA density profile.   The region enclosed within the
truncation boundary has to be intended as representative of
the quasi static halo interior, leaving aside the surrounding
material which is still infalling.   It is worth remembering
that the total mass can be finite even for an infinitely
extended configuration, provided the related density profile
is sufficiently steep e.g., $(\alpha,\beta,\gamma)=(1,4,1)$,
hereafter quoted as H density profile.

In dealing with numerical simulations of dark matter haloes,
it is usual to take into consideration spherically averaged
density profiles (e.g., Cole \& Lacey 1996; Navarro et al.
1997; Fukushige \& Makino 2001; Klypin et al. 2001), or in 
other terms spherical isopycnic (i.e. of
equal density) surfaces.   On the other hand, spin growth
by tidal interactions with neighbouring objects, in
expanding density perturbations, demands ellipsoidally
averaged density profiles (e.g., Doroshkevic 1970; White
1984; Maller et al. 2002; Jing \& Suto 2002), or in other 
terms ellipsoidal 
isopycnic surfaces.   As a best compromise between intrinsic 
simplicity and unavoidable necessity, our attention will be 
devoted to homeoidally striated configurations, i.e. the 
isopycnic surfaces are similar and similarly placed
ellipsoids.

Galaxies and cluster of galaxies may safely be idealized
as two subsystems which link only via gravitational 
interaction, in such a way that each component is
distorted by the tidal potential induced by the other.
Then the application of the virial theorem in tensor
form may be performed either to the whole system or
to each subsystem separately.   Towards this aim, the explicit
expression of the potential-energy tensors are needed.
Though some results are available in literature (e.g.,
Brosche et al. 1983;
Caimmi \& Secco 1992; Caimmi 1993, 1995), the related 
density profiles exhibit no central divergence, or
``cusp'', in contradiction with Eq.\,(\ref{eq:runi})
when $\gamma>0$.

The present attempt aims mainly to (i) formulate a 
general theory of homeoidally striated density profiles
with a central cusp; (ii) devote further investigation
to a few special cases which are consistent with the
results of both observations and simulations related
to galaxies and cluster of galaxies; (iii) apply to
galaxies, as the analogon to clusters
of galaxies performed in an earlier paper
(Caimmi, 2002).

The current paper is organized in the following way.   
The general theory of homeoidally striated, density 
profiles with a central cusp, is performed in 
Sect.\,\ref{teori}.
The general results are particularized to NFW, MOA, 
and H density profiles, which provide good fits to 
the results from both observations (e.g., Geller et al.
1999; Rines et al. 2001) and numerical simulations 
(e.g., Klypin et al. 2001), in Sect.\,\ref{spec}.   
An application to elliptical galaxies
is performed in Sect.\,\ref{apga}.    Finally, some
concluding remarks are reported in Sect.\,\ref{core}. 
Further details on fitting simulated and theoretical, 
self-similar, universal density profiles, are 
illustrated in the Appendix.   

\section{Homeoidally striated density profiles with a 
central cusp}\label{teori}

A general theory of homeoidally striated density profiles,
where no divergence in the density occurs, has been developed
in an earlier approach (Caimmi 1993, hereafter quoted as C93).
Here we take into consideration a more general approach,
where a central cusp may occur.    The isopycnic surfaces
are defined by the following law:
\begin{leftsubeqnarray}
\slabel{eq:rhoa}
&& \rho=\rho^\dagger f(\xi)~~;\quad f(1)=1~~; \\
\slabel{eq:rhob}
&& \xi^2=\sum_{\ell=1}^3\frac{x_\ell^2}{(a_\ell^\dagger)^
2}~~;\quad0\le\xi\le\Xi~~;
\label{seq:rho}
\end{leftsubeqnarray}
where $\rho^\dagger=\rho(1)$ , $a_\ell^\dagger$, are
the density and the semiaxes, respectively, of a
reference isopycnic surface, and $\Xi$ corresponds
to the truncation isopycnic surface, related to
semiaxes $a_\ell$.

Then the scaling density, $\rho^\dagger$, and the
scaling radius (related to the reference isopycnic 
surface), $r^\dagger$, 
correspond to a single boundary, which
allows the description of cuspy density profiles.
On the other hand, centrally regular density
profiles defined in C93 fail in this respect,
as the scaling density coincides with the central
density and the scaling radius attains the 
external surface.

The assumption that the system is homeoidally 
striated (Roberts 1962) implies the following
properties (e.g., C93):
\begin{leftsubeqnarray}
\slabel{eq:csia}
&& \xi=\frac{a_p^\prime}{a_p^\dagger}~~;\quad \diff
\xi=\frac{\diff a_p^\prime}{a_p^\dagger}~~;\quad
p=1,2,3~~; \\
\slabel{eq:csib}
&& \frac{\partial\xi}{\partial x_p}=\frac{x_p}
{\xi (a_p^\dagger)^2}~~;\quad p=1,2,3~~;\quad\sum_
{\ell=1}^3x_\ell\frac{\partial\xi}{\partial x_\ell}
=\xi~~;
\label{seq:csi}
\end{leftsubeqnarray}
where $a_p^\prime$ are the semiaxes
of the isopycnic surface under consideration.

The generic point on an isopycnic surface obeys
the equation:
\begin{equation}
\label{eq:ell1}
\frac{x_1^2}{a_1^{\prime2}}+\frac{x_2^2}{a_2^{\prime2}}+
\frac{x_3^2}{a_3^{\prime2}}=\frac{r^2\sin^2\theta\cos^2
\phi}{a_1^{\prime2}}+\frac{r^2\sin^2\theta\sin^2\phi}
{a_2^{\prime2}}+\frac{r^2\cos^2\theta}{a_3^{\prime2}}
=1~~;
\end{equation}
where ${\sf P}(x_1,x_2,x_3)\equiv{\sf P}(r,\theta,\phi)$.
Taking spherical coordinates into consideration,
Eq.\,(\ref{eq:ell1}) may be cast into the equivalent
form:
\begin{lefteqnarray}
\label{eq:ell2}
&& \frac{r^2}{(r^\dagger)^2}\frac{(r^\dagger)^2}{(a_1^\dagger)^2}\frac{(a_1^\dagger)^2}
{a_1^{\prime2}}\sin^2\theta\cos^2\phi+\frac{r^2}{(r^\dagger)^2}
\frac{(r^\dagger)^2}{(a_2^\dagger)^2}\frac{(a_2^\dagger)^2}{a_2^{\prime2}}\sin^2
\theta\sin^2\phi+\frac{r^2}{(r^\dagger)^2}\frac{(r^\dagger)^2}{(a_3^\dagger)^2}
\frac{(a_3^\dagger)^2}{a_3^{\prime2}}\cos^2\theta=\nonumber \\
&& \frac{r^2}{(r^\dagger)^2}\frac1{\xi^2}\left[\frac{(r^\dagger)^2}
{(a_1^\dagger)^2}\sin^2\theta\cos^2\phi+\frac{(r^\dagger)^2}{(a_2^\dagger)^2}
\sin^2\theta\sin^2\phi+\frac{(r^\dagger)^2}
{(a_3^\dagger)^2}\cos^2\theta\right]=1~~;
\end{lefteqnarray}
where $r^\dagger$ is the radial coordinate of the point, 
${\sf P}^\dagger(r^\dagger,\theta,\phi)$, on the reference isopycnic
surface.   Accordingly, the expression within
brackets in Eq.\,(\ref{eq:ell2}) makes the
particularization of Eq.\,(\ref{eq:ell1}) to the
reference isopycnic surface, and necessarily equals 
unity.   Then Eq.\,(\ref{eq:ell2}) reduces to:
\begin{equation}
\label{eq:csir}
\xi=\frac r{r^\dagger}~~;
\end{equation}
regardless of the radial direction.     The
particularization of Eq.\,(\ref{eq:csir}) to
the truncation boundary yields: 
\begin{equation}
\label{eq:CsiR}
\Xi=\frac R{r^\dagger}~~;
\end{equation}
regardless of the radial direction.

The above results may be summarized as follows.   
Given a homeoidally striated density profile, and
two generic points with coinciding angular coordinates,
${\sf P}(r,\theta,\phi)$ and ${\sf P}^\dagger(r^\dagger,
\theta,\phi)$, placed on two isopycnic surfaces,
$r(\theta,\phi)$ and $r^\dagger(\theta,\phi)$, the 
ratio of the radial coordinates, $\xi=r/r^\dagger$, does
not depend on the radial direction.   Then  Eqs.\,(\ref
{eq:csia}) follow from (\ref{eq:csir}) as special cases.

\subsection{Mass and inertia tensor}\label{inte}

The volume, the mass, and the inertia tensor of an
infinitely thin homeoid bounded by isopycnic
surfaces, may be expressed as (e.g., C93):
\begin{lefteqnarray}
\label{eq:dS}
&& \diff S^\prime=4\pi a_1^\dagger a_2^\dagger a_3^\dagger\xi^2\diff\xi=
3S^\dagger\xi^2\diff\xi~~; \\
\label{eq:dM}
&& \diff M^\prime=4\pi\rho^\dagger a_1^\dagger a_2^\dagger a_3^\dagger f(\xi)
\xi^2\diff\xi=3M^\dagger f(\xi)\xi^2\diff\xi~~; \\
\label{eq:dI}
&& \diff I_{pq}^\prime=\frac{4\pi}3\rho^\dagger\delta_{pq}
(a_p^\dagger)^3a_q^\dagger a_r^\dagger f(\xi)\xi^4\diff\xi=
\delta_{pq}M^\dagger(a_p^\dagger)^2f(\xi)\xi^4\diff\xi~~;
\end{lefteqnarray}
where $S^\dagger$ and $M^\dagger$ are the volume and mass,
respectively, of a homogeneous ellipsoid, bounded
by the reference isopycnic surface, and with the same
density, $\rho^\dagger$, as at the reference isopycnic
surface; and $\delta_{pq}$ is the Kronecker symbol.

Let us define, according to Roberts (1962):
\begin{equation}
\label{eq:Fc}
F(\xi)=2\int_\xi^\Xi f(\xi^\prime)\xi^\prime\diff
\xi^\prime~~;
\end{equation}
from which the following relations are easily
derived:
\begin{equation}
\label{eq:FC}
F(\Xi)=0~~;\quad\frac{\diff F}{\diff\xi}=-2\xi
f(\xi)~~;
\end{equation}
an integration by parts of Eq.\,(\ref{eq:Fc})
shows that:
\begin{equation}
\label{eq:inF}
\int_0^\Xi f(\xi)\xi^n\diff\xi=\frac{n-1}2\int_
0^\Xi F(\xi)\xi^{n-2}\diff\xi~~;\qquad n>1~~;
\end{equation}
which allows the calculation of the total mass as:
\begin{leftsubeqnarray}
\slabel{eq:Ma}
&& M=\nu_{mas}M^\dagger~~; \\
\slabel{eq:Mb}
&& \nu_{mas}=\frac32\int_0^\Xi F(\xi)\diff\xi~~; \\
\slabel{eq:Mc}
&& M^\dagger=\frac{4\pi}3\rho^\dagger a_1^\dagger a_2^\dagger a_3^\dagger~~;
\label{seq:M}
\end{leftsubeqnarray}
and the inertia tensor as:
\begin{leftsubeqnarray}
\slabel{eq:Ia}
&& I_{pq}=\delta_{pq}\nu_{inr}M^\dagger(a_p^\dagger)^2~~; \\
\slabel{eq:Ib}
&& \nu_{inr}=\frac32\int_0^\Xi F(\xi)\xi^2\diff\xi~~;
\label{seq:I}
\end{leftsubeqnarray}
where the coefficients, $\nu_{mas}$ and $\nu_{inr}$, are
shape-independent and may be conceived as profile
factors (C93).

The mass enclosed within a selected isopycnic
surface, $\rho=\rho^\dagger f(\xi)$, $0\le\xi\le\Xi$,
is obtained by integration of Eq.\,(\ref{eq:dM}).
The result is:
\begin{equation}
\label{eq:Mcsi}
M(\xi)=3M^\dagger\int_0^\xi f(\xi^\prime)\xi^{\prime2}
\diff\xi^\prime~~;
\end{equation}
and the related mean density is:
\begin{equation}
\label{eq:rhomc}
\bar{\rho}(\xi)=\frac{M(\xi)}{S(\xi)}=3\rho^\dagger
\frac1{\xi^3}\int_0^\xi f(\xi^\prime)\xi^{\prime2}
\diff\xi^\prime~~;
\end{equation}
where Eqs.\,(\ref{eq:dS}) and (\ref{eq:Mcsi})
have been taken into consideration.

\subsection{Tensor potential and self-energy tensor}
\label{sete}

The tensor potential induced by an infinitely
thin homeoid bounded by isopycnic surfaces, on
a generic point placed at its interior, is (e.g.,
C93):
\begin{equation}
\label{eq:dVpq}
\diff{\cal V}_{pq}^{(int)}=2\pi G\delta_{pq}\rho^\dagger
(a_p^\dagger)^2A_pf(\xi)\xi\diff\xi~~;
\end{equation}
where $G$ is the constant of gravitation, $p=1,2,3,$
$q=1,2,3,$ and $A_p$ are shape factors which, {\it 
ipso facto},
depend on the axis ratios only (e.g., Caimmi 1992;
therein defined as $\hat{\alpha}_p$).

The tensor potential induced by all the homeoids
enclosing a given point, on that point, is (e.g., C93):
\begin{equation}
\label{eq:Vpq}
{\cal V}_{pq}^{(int)}=\pi G\delta_{pq}\rho^\dagger
(a_p^\dagger)^2A_pF(\xi_p)~~;
\end{equation}
where $\xi_p$ denotes the isopycnic surface passing
through the point, and Eq.\,(\ref{eq:Fc}) has
been used.

The interaction-energy tensor related to an
infinitely thin homeoid bounded by isopycnic
surfaces, embedded into another infinitely
thin homeoid of the same kind, is (C93):
\begin{equation}
\label{eq:d2Wpq}
\diff^2(E_{int})_{pq}=-4\pi^2G\delta_{pq}(\rho^\dagger)^2
(a_p^\dagger)^3a_s^\dagger a_r^\dagger A_pf(\xi)\xi f(\xi
^\prime)\xi^{\prime2}\diff\xi\diff\xi^\prime~~;
\end{equation}
if the two boundaries coincide, $\xi$=$\xi^\prime$
and $\diff^2(E_{int})_{pq}(\xi$=$\xi^\prime)$=
$\diff^2(E_{sel})
_{pq}$, i.e. the interaction-energy tensor 
reduces to the self-energy tensor.    Bearing
in mind that Eq.\,(\ref{eq:d2Wpq}) remains
unchanged by replacing $\xi^\prime$ with $\xi$
and vice versa, the self-energy tensor of the 
whole mass distribution may be written as
$(E_{sel})_{pq}=2\int\int_S\diff^2(E_{int})_{pq}$; the 
result is (e.g., C93):
\begin{leftsubeqnarray}
\slabel{eq:Opqa}
&& (E_{sel})_{pq}=-\delta_{pq}\nu_{sel}\frac
{G(M^\dagger)^2}{a_1^\dagger}\epsilon_{p2}
\epsilon_{p3}A_p~~; \\
\slabel{eq:Opqb}
&& \nu_{sel}=\frac9{16}
\int_0^\Xi F^2(\xi)\diff\xi~~; \\
\slabel{eq:Opqc}
&& \epsilon_{pq}=\frac{a_p}{a_q}~~;
\label{seq:Opq}
\end{leftsubeqnarray}
where $\nu_{sel}$ is a profile factor, i.e.
shape-independent, and $\epsilon_{pq}$ are
axis ratios.  

Let ${\sf P}(r,\theta,\phi)$ be a generic point on
an isopycnic surface, $\rho=\rho^\dagger f(\xi)$,
$0\le\xi\le\Xi$.   A test particle at {\sf P} is
centrifugally supported with respect to the centre
of mass, provided the following relation holds:
\begin{equation}
\label{eq:eqc1}
\frac{GM(\xi)\eta(r,\theta,\phi)}{r^2}=\frac 
{v_{eq}^2(r,\theta,\phi)}r~~;
\end{equation}
where $\eta$ represents the deviation from the
spherical limit and/or the Roche-like (i.e. a 
mass point surrounded by a massless, ellipsoidal 
atmosphere) limit, and $v_{eq}$ is
the rotational velocity with respect to the
centre of mass.   Due to Newton's theorem,
a generic homeoid bounded by isopycnic surfaces,
related to $\xi^\prime>\xi$, $\xi^{\prime\prime}
>\xi$, exerts no action on {\sf P} (e.g.,
Chandrasekhar 1969, Chap.\,3, \S 17).   In this
respect, only the mass, $M(\xi)$, enclosed within
the isopycnic surface under consideration, is
effective.    

The particularization of Eq.\,(\ref{eq:eqc1})
to a point, ${\sf P}^\dagger(r^\dagger,\theta,\phi)$, on the
reference isopycnic surface, reads:
\begin{equation}
\label{eq:eqc2}
\frac{GM(1)\eta(r^\dagger,\theta,\phi)}{(r^\dagger)^2}=\frac
{v_{eq}^2(r^\dagger,\theta,\phi)}{r^\dagger}~~;
\end{equation}
and the combination of Eqs.\,(\ref{eq:csir}),
(\ref{eq:eqc1}) and (\ref{eq:eqc2}) yields:
\begin{equation}
\label{eq:v}
\frac{v_{eq}^2(r,\theta,\phi)}{v_{eq}^2(r^\dagger,
\theta,\phi)}=\frac1\xi\frac{M(\xi)}{M(1)}
\frac{\eta(r,\theta,\phi)}{\eta(r^\dagger,\theta,\phi)}~~;
\end{equation}
provided the point on the generic isopycnic
surface, ${\sf P}(r,\theta,\phi)$, and the
point on the reference isopycnic surface,
${\sf P}^\dagger(r^\dagger,\theta,\phi)$, are aligned
along the same radial direction.   

It is worth noting that the density profiles 
of interest, i.e. decreasing from the centre
to the boundary, lie between two extreme
situations: the homogeneous limit and the 
Roche-like limit, where the ratio, $\eta(r,
\theta,\phi)/\eta(r^\dagger,\theta,\phi)$, 
reduces to unity in both cases.

If the angular coordinates are specified, 
Eq.\,(\ref{eq:v}) takes the simpler form:
\begin{leftsubeqnarray}
\slabel{eq:vcsia}
&& \frac{v_{eq}^2(\xi)}{(v_{eq}^\dagger)^2}=\frac1\xi\frac
{M(\xi)}{M(1)}\frac{\eta(\xi)}{\eta^\dagger}~~; \\
\slabel{eq:vcsib}
&& v_{eq}^\dagger=v_{eq}(r^\dagger,\theta,\phi)~~;\qquad
\eta^\dagger=\eta(r^\dagger,\theta,\phi)~~;
\label{seq:vcsi}
\end{leftsubeqnarray}
where it is intended that the velocity profile
corresponds to the $(r^\dagger,\theta,\phi)$ direction.

\subsection{Angular-momentum vector and 
rotational-energy tensor}\label{arte}

In dealing with angular momentum and rotational
energy, the knowledge of the rotational velocity 
field is needed.   Let us define the angular-momentum
vector and the rotational-energy tensor, respectively,
as:
\begin{lefteqnarray}
\label{eq:Jpq}
&& J_p=\int\int\int_S\Omega_p\diff^3I_{\ell\ell}+\int\int
\int_S\Omega_p\diff^3I_{qq}\nonumber \\
&&-\int\int\int_S\Omega_\ell\diff^3I_{p\ell}+\int\int
\int_S\Omega_q\diff^3I_{pq}~~;\quad p\ne\ell~~;\quad 
q\ne\ell~~; \\ 
\label{eq:Erpq}
&& 2(E_{rot})_{pq}=\int\int\int_S\Omega_p\Omega_q
\diff^3I_{\ell\ell}-\int\int\int_S\Omega_\ell\Omega_p
\diff^3I_{q\ell}\nonumber \\
&&-\int\int\int_S\Omega_q\Omega_\ell\diff^3I_{\ell p}+
\int\int\int_S\Omega_\ell^2\diff^3I_{pq}~~;\quad 
p\ne\ell~~;\quad q\ne\ell~~;
\end{lefteqnarray}
where $\vec{\Omega}=(\Omega_1, \Omega_2, \Omega
_3)$ is the angular-velocity vector, and $\diff
^3I_{pq}$ is the inertia tensor, related to a 
generic, infinitesimal mass element.   

The symmetry of the tensors under consideration,
due to Eq.\,(\ref{eq:Ia}), together with the
conditions, $p\ne\ell$, $q\ne\ell$, implies the
following values for the index, $\ell$:
\begin{leftsubeqnarray}
\slabel{eq:rpqa}
&& \ell=1~~,\qquad p\vee q=2~~,\qquad p\ne1~\&~q\ne1~~; \\
\slabel{eq:rpqb}
&& \ell=2~~,\qquad p\vee q=3~~,\qquad p\ne2~\&~q\ne2~~; \\
\slabel{eq:rpqc}
&& \ell=3~~,\qquad p\vee q=1~~,\qquad p\ne3~\&~q\ne3~~;
\label{seq:rpq}
\end{leftsubeqnarray}
where $\vee$ and \& are the semantic symbol of 
disjunction, ``or'', and conjunction, ``and'',
respectively.

The preservation of ellipsoidal shape imposes
severe constraints on the rotational velocity 
field.   Leaving an exhaustive investigation
to more refined approaches, our attention will 
be limited to 
a restricted number of special cases, namely: (i)
rigid rotation about a principal axis, and (ii)
differential rotation about a symmetry axis, in
connection with a particular class of velocity
distributions.    By ``rotation'' it is intended,
of course, circular rotation.
Accordingly, $\vec{\Omega}=
(\delta_{1s}\Omega_1, \delta_{2s}\Omega_2, \delta_
{3s}\Omega_3)$ provided $x_s$ is assumed as
rotation axis.   

The angular momentum and the
rotational energy, related to axis $x_s$, are:
\begin{lefteqnarray}
\label{eq:Jp}
&& J(s)=J_s=\int\int\int_S\Omega_s\diff^3I_{pp}+
\int\int\int_S\Omega_s\diff^3I_{qq}~~;\quad p\ne
q\ne s~~; \\ 
\label{eq:Erp}
&& E_{rot}(s)=(E_{rot})_s=(E_{rot})_{pp}+(E_
{rot})_{qq}\nonumber \\
&& =\int\int\int_S\Omega_s^2\diff^3I_{pp}+
\int\int\int_S\Omega_s^2\diff^3I_{qq}~~;\quad 
p\ne q\ne s~~;
\end{lefteqnarray}
where $s=3$ is usually assumed, and then $p=1$,
$q=2$, or vice versa.   For the sake of brevity, from
this point on $J_s$ and $(E_{rot})_s$ shall be
denoted as $J$ and $E_{rot}$, respectively.

In the special case of rigid rotation about a 
principal axis, 
Eq.\,(\ref{eq:Erpq}), by use of (\ref{eq:Ia}),
reads:
\begin{lefteqnarray}
\label{eq:Erpqr}
&& (E_{rot})_{pq}=\frac12I_{pq}\Omega_r^2=
\frac12\delta_{pq}(1-\delta_{pr})\nu_{inr}M^\dagger(a_p^
\dagger)^2\Omega_r^2=\frac12\delta_{pq}(1-\delta_{pr})\nu_
{inr}M^\dagger(v_{rot}^\dagger)_p^2~~;
\end{lefteqnarray}
and Eqs.\,(\ref{eq:Jp}), (\ref{eq:Erp}), taking
$s=3$, read:
\begin{lefteqnarray}
\label{eq:Jr}
&& J=(I_{11}+I_{22})\Omega=\nu_{inr}M^\dagger a_1^\dagger
(1+\epsilon_{21}^2)v_{rot}^\dagger~~; \\ 
\label{eq:Err}
&& E_{rot}=\frac12\nu_{inr}M^\dagger(1+\epsilon_{21}^2)
(v_{rot}^\dagger)^2~~;
\end{lefteqnarray}
where $\vec{v}_{rot}=\vec{\Omega}\times\vec{r}$ and
$v_{rot}^\dagger=(v_{rot}^\dagger)_1$ is the rotational 
velocity at the end of the major semiaxis, $a_1^\dagger$,
with regard to the reference isopycnic surface.

In the special case of differential rotation about
a symmetry axis, our attention shall be restricted
to (rotational) velocity distributions which satisfy
the counterpart of Eq.\,(\ref{eq:csir}), namely:
\begin{equation}
\label{eq:vrot}
\frac{v_{rot}(r,\theta)}{v_{rot}(r^\dagger,\theta)}=
\frac{v_{rot}(a^\prime,0)}{v_{rot}(a^\dagger,0)}~~;
\end{equation}
where $(r,\theta)$, $(r^\dagger,\theta)$, represent a
point on a generic and reference isopycnic surface, 
respectively, along the same radial direction, 
and $(a^\prime,0)$, $(a^\dagger,0)$, represent
the end of the corresponding, major semiaxes.
Due to Eqs.\,(\ref{eq:csia}) and (\ref{eq:csir}),
Eq.\,(\ref{eq:vrot}) may be written under the
equivalent form:
\begin{equation}
\label{eq:omega}
\frac{\Omega(r,\theta)}{\Omega(r^\dagger,\theta)}=
\frac{\Omega(a^\prime,0)}{\Omega(a^\dagger,0)}~~;
\end{equation}
where rotational velocity is replaced by angular
velocity.   It is worth noting that, in 
particular, either rotational or angular
velocity is allowed to be constant everywhere.

Both rotational and angular velocity are 
independent of the longitudinal angle, $\phi$,
according to Eqs.\,(\ref{eq:vrot}) and (\ref
{eq:omega}), respectively.   It follows that
sections of infinitely thin, homogeneous
homeoids, by infinitely thin layers normal
to the symmetry axis, do rotate rigidly.
The related volume, inertia tensor, 
angular-momentum vector, rotational-energy 
tensor, are (e.g., Caimmi \& Secco 1993):
\begin{lefteqnarray}
\label{eq:d2S}
&& \diff^2S=\frac{2\pi\epsilon^3(a^\dagger)^3\xi^2\sin\theta
\diff\xi\diff\theta}{(\cos^2\theta+\epsilon^2\sin^2
\theta)^{3/2}}~~;
\end{lefteqnarray}
\begin{leftsubeqnarray}
\slabel{eq:d2I1}
&& \diff^2I_{1q}=\diff^2I_{2q}=\frac34\delta_{1q}
M^\dagger (a^\dagger)^2\epsilon^4f(\xi)\xi^4\frac{\sin^3\theta
\diff\xi\diff\theta}{(\cos^2\theta+\epsilon^2\sin^2
\theta)^{5/2}}~~; \\ 
\slabel{eq:d2I3}
&& \diff^2I_{3q}=\frac32\delta_{3q}M^\dagger(a^\dagger)
^2\epsilon^4f(\xi)\xi^4\frac{\sin\theta\cos^2\theta
\diff\xi\diff\theta}{(\cos^2\theta+\epsilon^2\sin^2
\theta)^{5/2}}~~;
\label{seq:d2I}
\end{leftsubeqnarray}
\begin{lefteqnarray}
\label{eq:d2Jpq}
&& \diff^2J=\Omega(\xi,\theta)\left[\diff^2I_{11}+
\diff^2I_{22}\right]~~; \\
\label{eq:d2Erpq}
&& \diff^2(E_{rot})_{pq}=\frac12\delta_{pq}(1-\delta_{p3})
\Omega^2(\xi,\theta)\diff^2I_{pq}~~;
\end{lefteqnarray}
where, due to axial symmetry, $a_1^\dagger=a_2^\dagger=a^\dagger$,
$\epsilon_{21}=1$, $\epsilon_{31}=\epsilon$, and 
$\theta$ is the azimuthal angle.   The angular-momentum
vector and the rotational-energy tensor, by use of 
Eqs.\,(\ref{eq:Mc}) and (\ref{eq:omega})-(\ref
{eq:d2Erpq}), take the equivalent form:
\begin{lefteqnarray}
\label{eq:ddJpq}
&& \diff^2J=
\frac32M^\dagger (a^\dagger)^2\epsilon^4f(\xi)\xi^4
\Omega(\xi,0)\frac{\Omega(1,\theta)}{\Omega(1,0)}
\frac{\sin^3\theta\diff\xi\diff\theta}{(\cos^2\theta+
\epsilon^2\sin^2\theta)^{5/2}}~~; \\
\label{eq:ddErpq}
&& \diff^2(E_{rot})_{pq}=\frac38\delta_{pq}(1-\delta_{p3})
M^\dagger (a^\dagger)^2\epsilon^4f(\xi)\xi^4\Omega^2(\xi,0)
\frac{\Omega^2(1,\theta)}{\Omega^2(1,0)}\frac
{\sin^3\theta\diff\xi\diff\theta}{(\cos^2\theta+
\epsilon^2\sin^2\theta)^{5/2}}~~;
\end{lefteqnarray}
where, due to Eqs.\,(\ref{eq:vrot}), (\ref{eq:omega}), 
the ratios, $v_{rot}(1,\theta)/v_{rot}(1,0)$, $\Omega
(1,\theta)/\Omega(1,0)$, only depend on the azimuthal angle.

An integration over the whole, homogeneous, infinitely
thin homeoid yields:
\begin{lefteqnarray}
\label{eq:dJpqa}
&& \diff J=2
\eta_{anm}M^\dagger(a^\dagger)^2\Omega(\xi,0)f(\xi)\xi^4\diff\xi~~; \\
\label{eq:dJpqb}
&& \eta_{anm}=\frac34\epsilon^4\int_{-\pi/2}^{+\pi/2}
\frac{\Omega(1,\theta)}{\Omega(1,0)}\frac
{\sin^3\theta\diff\theta}{(\cos^2\theta+
\epsilon^2\sin^2\theta)^{5/2}}~~;
\end{lefteqnarray}
\begin{lefteqnarray}
\label{eq:dErpqa}
&& \diff(E_{rot})_{pq}=\delta_{pq}(1-\delta_{p3})
\eta_{rot}M^\dagger (a^\dagger)^2\Omega^2(\xi,0)f(\xi)\xi^4\diff
\xi~~; \\
\label{eq:dErpqb}
&& \eta_{rot}=\frac38\epsilon^4\int_{-\pi/2}^{+\pi/2}
\frac{\Omega^2(1,\theta)}{\Omega^2(1,0)}\frac
{\sin^3\theta\diff\theta}{(\cos^2\theta+
\epsilon^2\sin^2\theta)^{5/2}}~~;
\end{lefteqnarray}
finally, an integration over the whole system yields, 
for the angular-momentum vector:
\begin{leftsubeqnarray}
\slabel{eq:Jpqa}
&& J=2
\eta_{anm}\nu_{anm}M^\dagger v_{rot}^\dagger a^\dagger~~; \\
\slabel{eq:Jpqb}
&& \nu_{anm}=
\int_0^\Xi\frac{\Omega
(\xi,0)}{\Omega(1,0)}f(\xi)\xi^4\diff\xi~~;
\label{seq:Jpq}
\end{leftsubeqnarray}
and for the rotational-energy tensor:
\begin{leftsubeqnarray}
\slabel{eq:Erpqa}
&& (E_{rot})_{pq}=\delta_{pq}(1-\delta_{p3})
\eta_{rot}\nu_{rot}M^\dagger(v_{rot}^\dagger)^2~~; \\
\slabel{eq:Erpqb}
&& \nu_{rot}=
\int_0^\Xi\frac{\Omega
^2(\xi,0)}{\Omega^2(1,0)}f(\xi)\xi^4\diff\xi~~;
\label{seq:Erpq}
\end{leftsubeqnarray}
it is apparent that $\eta_{anm}$, $\eta_{rot}$, depend
on the integration over the boundary and may be 
thought of as shape factors, even if they may also 
be shape-independent, whereas $\nu_{anm}$, $\nu_{rot}$, 
are genuine profile factors.

The special case of rigid rotation, implies the
relation:
\begin{equation}
\label{eq:omc}
\frac{\Omega(\xi,\theta)}{\Omega(\xi,0)}=1~~;
\end{equation}
the special case of constant rotational velocity
on the equatorial plane, implies the relation:
\begin{equation}
\label{eq:omv}
\frac{\Omega(\xi,0)}{\Omega(1,0)}=\frac{a^\dagger}
{a^\prime}=\frac1\xi~~;
\end{equation}
the special case of constant rotational velocity
everywhere, implies the relation:
\begin{equation}
\label{eq:vrc}
\frac{\Omega(\xi,\theta)}{\Omega(\xi,0)}=\frac{a^
\prime}{r\sin\theta}=\frac{(\cos^2\theta+\epsilon^2
\sin^2\theta)^{1/2}}{\epsilon\sin\theta}~~;
\end{equation}
the corresponding values of the shape parameters,
$\eta_{anm}$, $\eta_{rot}$, and profile parameters, 
$\nu_{anm}$, $\nu_{rot}$, are listed in Table~\ref
{t:profo}.
\begin{table}
\begin{tabular}{cccc}
\hline
\hline
\multicolumn{1}{c}{case} &
\multicolumn{1}{c}{RR}  &
\multicolumn{1}{c}{RC}  &
\multicolumn{1}{c}{CV} \\
\hline
isopycnic surface & $\frac{\Omega(\xi,\theta)}{\Omega(\xi,0)}=1$ 
& $\frac{\Omega(\xi,\theta)}{\Omega(\xi,0)}=1$ & $\frac{v_{rot}(\xi,
\theta)}{v_{rot}(\xi,0)}=1$ \\
equatorial plane & $\frac{\Omega(\xi,0)}{\Omega(1,0)}=1$ 
& $\frac{v_{rot}(\xi,0)}{v_{rot}(1,0)}=1$ & $\frac{v_{rot}(\xi,
0)}{v_{rot}(1,0)}=1$ \\
\hline
$\eta_{anm}$ & 1 & 1 & 3$\pi$/8 \\
$\eta_{rot}$ & 1/2 & 1/2 & 3/4 \\
$\nu_{anm}$ & $\nu_{inr}$ &  $\int_0^\Xi f(\xi)\xi^3\diff\xi$ & 
$\int_0^\Xi f(\xi)\xi^3\diff\xi$  \\
$\nu_{rot}$ & $\nu_{inr}$ & $\nu_{mas}/3$ & $\nu_{mas}/3$ \\
\hline\hline
 & & & \\
\end{tabular}
\caption{Values of the shape
parameters, $\eta_{anm}$, $\eta_{rot}$, and profile
parameters, $\nu_{anm}$, $\nu_{rot}$, related to (i)
homeoidally striated ellipsoids in rigid rotation
about a principal axis (RR), and (ii) homeoidally
striated spheroids in differential rotation about
the symmetry axis where the isopycnic surfaces are:
either in rigid rotation and the rotational
velocity on the equatorial plane is constant (RC),
or in cylindric rotation and the rotational velocity 
is constant everywhere(CV).}
\label{t:profo}
\end{table}

\subsection{Other potential-energy tensors for
subsystems with similar boundaries}\label{oite}

The above results are related to a single
subsystem: strictly speaking, all the
quantities discussed in the last Section
should be labelled by the index,
$u$, in connection with $u$th subsystem
(e.g., Caimmi \& Secco 1992), but it has
been omitted for the sake of clarity.   On
the other hand, the formulation of the
interaction-energy tensor and the
tidal-energy tensor necessarily involves
at least two components.   The related
calculations turn out to be very difficult
in the general case, and for this reason
our attention shall be limited to subsystems
with similar boundaries, and homeoidally
striated density profiles with a central
cusp.

Using a similar procedure as in C93,
where no (central) cusp occurs in density
profiles, but bearing in mind that the 
scaled density and the scaled radius have 
a different definition, yields the 
expression of the interaction-energy tensors:
\begin{leftsubeqnarray}
\slabel{eq:Wija}
&& [(E_{ij})_{int}]_{pq}=-\delta_{pq}
\frac{G(M_i^\dagger)^2}{(a_i^\dagger)_1}(\nu_{ij})_{int}\epsilon_
{p2}\epsilon_{p3}A_p~~; \\
\slabel{eq:Wijb}
&& (\nu_{ij})_{int}=-\frac9{16}m^\dagger
\left[w^{(int)}(\Xi_i/y^\dagger)+w^{(ext)}(\Xi_i/y^
\dagger)\right]~~; \\
\slabel{eq:Wijc}
&& [(E_{ji})_{int}]_{pq}=[(E_{ij})_{int}]_{pq}~~;
\label{seq:Wij}
\end{leftsubeqnarray}
the expression of the tidal-energy tensors:
\begin{leftsubeqnarray}
\slabel{eq:Vija}
&& [(E_{ij})_{tid}]_{pq}=-\delta_{pq}\frac{G(M_i^\dagger)^2}
{(a_i^\dagger)_1}(\nu_{ij})_{tid}\epsilon_{p2}\epsilon_
{p3}A_p~~; \\
\slabel{eq:Vijb}
&& (\nu_{ij})_{tid}=-\frac98m^\dagger
w^{(ext)}(\Xi_i/y^\dagger)~~; \\
\slabel{eq:Vijc}
&& [(E_{ji})_{tid}]_{pq}=-\delta_{pq}
\frac{G(M_j^\dagger)^2}{(a_j^\dagger)_1}(\nu_{ji})_{tid}\epsilon_
{p2}\epsilon_{p3}A_p~~; \\
\slabel{eq:Vijd}
&& (\nu_{ji})_{tid}=-\frac98\frac{y^\dagger}{m^\dagger}
w^{(int)}(\Xi_i/y^\dagger)~~;
\label{seq:Vij}
\end{leftsubeqnarray}
and the expression of the residual-energy tensors:
\begin{leftsubeqnarray}
\slabel{eq:Qija}
&& [(E_{ij})_{res}]_{pq}=-\delta_{pq}
\frac{G(M_i^\dagger)^2}{(a_i^\dagger)_1}(\nu_{ij})_{res}
\epsilon_{p2}\epsilon_{p3}A_p~~; \\
\slabel{eq:Qijb}
&& (\nu_{ij})_{res}=-\frac9{16}m^\dagger
\left[w^{(ext)}(\Xi_i/y^\dagger)-w^{(int)}(\Xi_i/y^
\dagger)\right]~~; \\
\slabel{eq:Qijc}
&& [(E_{ji})_{res}]_{pq}=-[(E_{ij})_{res}]_{pq}~~;
\label{seq:Qij}
\end{leftsubeqnarray}
where the functions, $w^{(int)}$ and $w^{(ext)}$,
are defined as:
\begin{leftsubeqnarray}
\slabel{eq:wiea}
&& w^{(int)}(\eta)=\int_0^\eta F_j(\xi_j)\frac
{\diff F_i}{\diff\xi_j}\xi_j\diff\xi_j~~; \\
\slabel{eq:wieb}
&& w^{(ext)}(\eta)=\int_0^\eta F_i(\xi_i)\frac
{\diff F_j}{\diff\xi_j}\xi_j\diff\xi_j~~;
\label{seq:wie}
\end{leftsubeqnarray}
the parameters, $m^\dagger$ and $y^\dagger$,
are defined as:
\begin{equation}
\label{eq:mdyd}
m^\dagger=\frac{M_j^\dagger}{M_i^\dagger}~~;\qquad
y^\dagger=\frac{r_j^\dagger}{r_i^\dagger}~~;
\end{equation}
and the combination of Eqs.\,(\ref{eq:CsiR}) and
(\ref{eq:mdyd}) yields:
\begin{equation}
\label{eq:Csiy}
\frac{\Xi_j}{\Xi_i}=\frac y{y^\dagger}~~;\qquad
y=\frac{R_j}{R_i}~~;
\end{equation}
finally, the virial-energy tensors (e.g., Caimmi
\& Secco 2002), $[(E_{uv})_{vir}]_{pq}=[(E_u)_{sel}]
_{pq}+[(E_{uv})_{tid}]_{pq}$, take the expression:
\begin{leftsubeqnarray}
\slabel{eq:Uija}
&&[(E_{uv})_{vir}]_{pq}=-\delta_{pq}\frac{G(M_u^
\dagger)^2}{(a_u^\dagger)_1}(\nu_{uv})_{vir}
\epsilon_{p2}\epsilon_{p3}A_p~~; \\
\slabel{eq:Uijb}
&& (\nu_{uv})_{vir}=(\nu_u)_{sel}+(\nu_{uv})_{tid}
~~;\qquad u=i,j~~;\qquad v=j,i~~;
\label{seq:Uij}
\end{leftsubeqnarray}
in conclusion, Eqs.\,(\ref{seq:Wij})-(\ref{seq:Uij})
allow the calculation of potential-energy tensors
for homeoidally striated density profiles with a
central cusp, related to two subsystems with
similar and similarly placed boundaries.

\section{Special cases}\label{spec}

The physical parameters related to the density profiles
expressed by Eq.\,(\ref{eq:runi}), and investigated
in Sect.\,2, may be explicitly calculated, provided
the exponents, $(\alpha, \beta, \gamma)$, are
specified.    Our attention shall be limited to
two special cases, which are consistent with
high-resolution simulations of cold dark matter
haloes (e.g., Klypin et al. 2001): NFW, first 
recognized by Navarro et al. (1995, 1996, 1997), 
and MOA, first
recognized by Moore et al. (1998, 1999).   As
far as we know, no method has been clearly outlined
for fitting simulated and theoretical, 
self-similar, universal density profiles.
A possible procedure is sketched out in 
Appendix\,\ref{espo} for a single matter
distribution which obeys Eq.\,(\ref{eq:runi}),
and in Appendix\,\ref{proFM} for two
different mass distributions which obey
Eq.\,(\ref{eq:runi}).   
In addition, a third density profile will be 
taken into consideration, which is consistent 
(together with NFW) with observations (e.g., 
Geller et al. 1999; Rines et al. 2001; 
Holley-Bockelmann et al. 2001): H, 
first proposed by Hernquist (1990).

With regard to the shape parameters, 
$\eta_{anm}$, $\eta_{rot}$, and the 
profile parameters, $\nu_{anm}$, $\nu_{rot}$,
it is worth remembering that they
are listed in Table~\ref{t:profo} for the
velocity profiles discussed in Sect.\,\ref{teori},
where differential rotation corresponds to
axisymmetric configurations with constant 
rotational velocity,
either on the equatorial plane or everywhere.

\subsection{The NFW density profile}\label{NFW}

The NFW density profile corresponds to the choice
of exponents $(\alpha, \beta, \gamma)=(1,3,1)$ in
Eq.\,(\ref{eq:runi}).    Accordingly, Eq.\,(\ref
{eq:rhoa}) reads:
\begin{equation}
\label{eq:rhoN}
\rho(\xi)=\rho^\dagger f(\xi)~~;\quad f(\xi)=\frac4{\xi
(1+\xi)^2}~~;
\end{equation}
and Eq.\,(\ref{eq:Fc}), after integration, takes
the explicit expression:
\begin{equation}
\label{eq:FN}
F(\xi)=\frac8{1+\xi}-\frac8{1+\Xi}~~;
\end{equation}
where the truncated, scaled radius, defined
by Eq.\,(\ref{eq:CsiR}), is the analogon of
the concentration, $c$, defined in Navarro
et al. (1997).

The combination of Eqs.\,(\ref{eq:inF}),  
(\ref{eq:Mcsi}), (\ref{eq:rhomc}),
(\ref{eq:rhoN}), and (\ref{eq:FN}) yields:
\begin{lefteqnarray}
\label{eq:McN}
&& M(\xi)=12M^\dagger\left[\ln(1+\xi)-\frac\xi{1+\xi}
\right]~~; \\
\label{eq:rmcN}
&& \bar{\rho}(\xi)=12\rho^\dagger\frac1{\xi^3}\left[
\ln(1+\xi)-\frac\xi{1+\xi}\right]~~;
\end{lefteqnarray}
which represent the mass and mean density
enclosed within a generic, isopycnic surface.
It is worth noting that the profile parameter, 
$\nu_{\bar{\rho}}=\bar{\rho}(\Xi)/\rho^\dagger$, 
is linked to the dimensionless density, $\delta_
c$, defined in Navarro et al. (1997), via the 
relation $\delta_c=800/\nu_{\bar{\rho}}$.

The profile parameters, $\nu_{mas}$, $\nu_{inr}$, 
and $\nu_{sel}$, due to Eqs.\,(\ref{eq:Mb}), 
(\ref{eq:Ib}), (\ref{eq:Opqb}), and (\ref{eq:FN}),
after integration, take the explicit expressions:
\begin{lefteqnarray}
\label{eq:nuMN}
&& \nu_{mas}=12\left[\ln(1+\Xi)-\frac\Xi{1+\Xi}
\right]~~; \\
\label{eq:nuIN}
&& \nu_{inr}=\frac43\left[\frac{\Xi(\Xi^2-3\Xi-6)}
{1+\Xi}+6\ln(1+\Xi)\right]~~; \\
\label{eq:nuON}
&& \nu_{sel}=36\frac{\Xi(2+\Xi)-2(1+\Xi)
\ln(1+\Xi)}{(1+\Xi)^2}~~;
\end{lefteqnarray}
the velocity profile related to centrifugal 
support along a radial direction, by use of
Eqs.\,(\ref{eq:vcsia}) and (\ref
{eq:McN}), is:
\begin{equation}
\label{eq:veqN}
v_{eq}(\xi)=v_{eq}^\dagger\left[\frac1\xi\frac{\ln
(1+\xi)-\xi/(1+\xi)}{\ln2-1/2}\right]^{1/2}
\left[\frac{\eta(\xi)}{\eta^\dagger}\right]^{1/2}~~;
\end{equation}
and the profile parameter, $\nu_{anm}$, due to 
Eq.\,(\ref{eq:rhoN}), after integration, takes 
the explicit expression:
\begin{equation}
\label{eq:nuJN}
\nu_{anm}=4\Xi\left[\frac{\Xi(2+\Xi)}
{1+\Xi}-2\ln(1+\Xi)\right]~~;
\end{equation}
which completes the specification of the
physical parameters of interest, for the
case under discussion.

For further analysis of the properties of mass
distributions with NFW density profiles (which 
is outside the aim of the current paper) see 
e.g., Lokas \& Mamon (2001).

\subsection{The MOA density profile}\label{MOA}

The MOA density profile corresponds to the choice
of exponents $(\alpha, \beta, \gamma)=(3/2,3,3/2)$ 
in Eq.\,(\ref{eq:runi}).    Accordingly, Eq.\,(\ref
{eq:rhoa}) reads:
\begin{equation}
\label{eq:rhoM}
\rho(\xi)=\rho^\dagger f(\xi)~~;\quad f(\xi)=\frac2{\xi^
{3/2}(1+\xi^{3/2})}~~;
\end{equation}
and Eq.\,(\ref{eq:Fc}), after integration, takes
the explicit expression:
\begin{lefteqnarray}
\label{eq:FM}
&& F(\xi)=\frac43\ln\frac{(1+\Xi^{1/2})^2}{1-\Xi^{1/2}
+\Xi}+\frac8{\sqrt{3}}\arctg\frac{2\Xi^{1/2}-1}
{\sqrt{3}} \nonumber \\
&& \phantom{F(\xi)=\hspace{1mm}}-\frac43\ln\frac
{(1+\xi^{1/2})^2}{1-\xi^{1/2}+\xi}-\frac8{\sqrt{3}}
\arctg\frac{2\xi^{1/2}-1}{\sqrt{3}}~~;
\end{lefteqnarray}
the combination of Eqs.\,(\ref{eq:inF}),  
(\ref{eq:Mcsi}), (\ref{eq:rhomc}),
(\ref{eq:rhoM}), and (\ref{eq:FM}) yields:
\begin{lefteqnarray}
\label{eq:McM}
&& M(\xi)=4M^\dagger\ln(1+\xi^{3/2})~~; \\
\label{eq:rmcM}
&& \bar{\rho}(\xi)=4\rho^\dagger\frac1{\xi^3}
\ln(1+\xi^{3/2})~~;
\end{lefteqnarray}
which represent the mass and mean density
enclosed within a generic, isopycnic surface.

The profile parameters, $\nu_{mas}$ and $\nu_{inr}$, 
due to Eqs.\,(\ref{eq:Mb}), (\ref{eq:Ib}), 
(\ref{eq:Opqb}), and (\ref{eq:FM}),
after integration, take the explicit expressions:
\begin{lefteqnarray}
\label{eq:nuMM}
&& \nu_{mas}=4\ln(1+\Xi^{3/2})~~; \\
\label{eq:nuIM}
&& \nu_{inr}=-4\Xi^{1/2}+
\Xi^2+\frac4{\sqrt{3}}\arctg\frac{2\Xi^{1/2}-1}
{\sqrt{3}} \nonumber \\
&& \phantom{\nu_{inr}=~}+\frac43\ln(1+\Xi^{1/2})-\frac23
\ln(1-\Xi^{1/2}+\Xi)+\frac{2\pi}{3\sqrt{3}}~~;
\end{lefteqnarray}
the profile parameter, $\nu_{sel}$, could also be
analytically integrated%
\footnote{By visiting the internet site:
``HTTP://INTEGRALS.WOLFRAM.COM/INDEX.CGI''.},
but the result is exceedingly long and, in practice,
unserviceable.   For this reason, it has to be
calculated numerically.

The velocity profile related to centrifugal 
support along a radial direction, by use of
Eqs.\,(\ref{eq:vcsia}) and (\ref{eq:McM}), is:
\begin{equation}
\label{eq:veqM}
v_{eq}(\xi)=v_{eq}^\dagger\left[\frac1\xi\frac{\ln
(1+\xi^{3/2})}{\ln2}\right]^{1/2}\left[\frac
{\eta(\xi)}{\eta^\dagger}\right]^{1/2}~~;
\end{equation}
and the profile parameter, $\nu_{anm}$, due to 
Eq.\,(\ref{eq:rhoM}), after integration, takes 
the explicit expression:
\begin{lefteqnarray}
\label{eq:nuJM}
&& \nu_{anm}=2\Xi\left[\Xi-\frac2{\sqrt{3}}
\arctg\frac{2\Xi^{1/2}-1}{\sqrt{3}}+\frac23\ln(1+
\Xi^{1/2})\right. \nonumber \\
&& \phantom{\nu_{anm}=}\left.-\frac13\ln(1-\Xi^{1/2}+\Xi)-
\frac\pi{3\sqrt{3}}\right]~~;
\end{lefteqnarray}
which completes the specification of the
physical parameters of interest, for the
case under discussion.

\subsection{The H density profile}\label{H}

The H density profile corresponds to the choice
of exponents $(\alpha, \beta, \gamma)=(1,4,1)$ in
Eq.\,(\ref{eq:runi}).    Accordingly, Eq.\,(\ref
{eq:rhoa}) reads:
\begin{equation}
\label{eq:rhoH}
\rho(\xi)=\rho^\dagger f(\xi)~~;\quad f(\xi)=\frac8{\xi
(1+\xi)^3}~~;
\end{equation}
and Eq.\,(\ref{eq:Fc}), after integration, takes
the explicit expression:
\begin{equation}
\label{eq:FH}
F(\xi)=\frac8{(1+\xi)^2}-\frac8{(1+\Xi)^2}~~;
\end{equation}
the combination of Eqs.\,(\ref{eq:inF}),  
(\ref{eq:Mcsi}), (\ref{eq:rhomc}),
(\ref{eq:rhoH}), and (\ref{eq:FH}) yields:
\begin{lefteqnarray}
\label{eq:McH}
&& M(\xi)=12M^\dagger\frac{\xi^2}{(1+\xi)^2}~~; \\
\label{eq:rmcH}
&& \bar{\rho}(\xi)=12\rho^\dagger\frac1{\xi^3}\frac{\xi^2}
{(1+\xi)^2}~~;
\end{lefteqnarray}
which represent the mass and mean density
enclosed within a generic, isopycnic surface.

The profile parameters, $\nu_{mas}$, $\nu_{inr}$, 
and $\nu_{sel}$, due to Eqs.\,(\ref{eq:Mb}), 
(\ref{eq:Ib}), (\ref{eq:Opqb}), and (\ref{eq:FH}),
after integration, take the explicit expressions:
\begin{lefteqnarray}
\label{eq:nuMH}
&& \nu_{mas}=12\frac{\Xi^2}{(1+\Xi)^2}~~; \\
\label{eq:nuIH}
&& \nu_{inr}=12\left[\frac{\Xi(2+\Xi)}{1+\Xi}
-2\ln(1+\Xi)\right]~~; \\
\label{eq:nuOH}
&& \nu_{sel}=12\frac{\Xi^3(4+\Xi)}{(1+\Xi)^4}~~;
\end{lefteqnarray}
the velocity profile related to centrifugal 
support along a radial direction, by use of
Eqs.\,(\ref{eq:vcsia}) and (\ref{eq:McH}), is:
\begin{equation}
\label{eq:veqH}
v_{eq}(\xi)=v_{eq}^\dagger\frac{2\xi^{1/2}}{1+\xi}
\left[\frac{\eta(\xi)}{\eta^\dagger}\right]^{1/2}~~;
\end{equation}
and the profile parameter, $\nu_{anm}$, due to 
Eq.\,(\ref{eq:rhoH}), after integration, takes 
the explicit expression:
\begin{equation}
\label{eq:nuJH}
\nu_{anm}=4\Xi\left[-\frac{\Xi
(2+3\Xi)}{(1+\Xi)^2}+2\ln(1+\Xi)\right]~~;
\end{equation}
which completes the specification of the
physical parameters of interest, for the
case under discussion.

\subsection{Comparison with observations and 
simulations}\label{cNM}

Self-similar, universal density profiles are
characterized by two independent parameters,
i.e. a scaling density, $\rho^\dagger$, and a
scaling radius, $r^\dagger$, according to Eq.\,(\ref
{eq:runi}).   In some (e.g., NFW and MOA 
density profiles) but not all (e.g., H density
profile) cases, the density must be null outside a
truncation radius, to avoid an infinite, total 
mass.   The truncation radius
cannot exceed the tidal radius, due to the 
presence of neighbouring objects, and may
safely be put equal to the virial radius
(e.g., Cole \& Lacey 1996; Navarro et al.
1997; Fukushige \& Makino 2001; Klypin et
al. 2001).    In fact, the inner, denser
regions of density perturbations collapse
and virialize first, while the outer, less
dense regions are still expanding or 
falling in.

High-resolution, dark matter halo 
simulations from hierarchical
clustering in either CDM or $\Lambda$CDM
scenarios, are known to be consistent 
with both NFW and MOA density profiles (e.g., 
Fukushige \& Makino 2001; Klypin et al. 2001).
Then the particularization of homeoidally
striated, density profiles with a central 
cusp discussed in Sect.\,\ref{teori}, to the
special cases NFW and MOA, makes a useful
tool for investigating some basic properties 
of large-scale celestial objects, such as
galaxies or clusters of galaxies.

With regard to elliptical galaxies, some
properties of the stellar subsystem are
reproduced, to an acceptable extent, by
H density profiles, in agreement with 
the observations (e.g., Hernquist 1990;
Holley-Bockelmann et al. 2001).   It is
worth remembering that H density profiles
closely approximate the de Vaucouleurs
$r^{1/4}$ law for elliptical galaxies.
In addition, recent observations (Geller
et al. 1999; Rines et al. 2001) allow
for the Coma cluster of galaxies both
NFW and H density profiles, to represent
the whole mass distribution within about 
10$h^{-1}$Mpc, for a virial radius equal
to 1.5$h^{-1}$Mpc.   The results of the current paper
were applied to the Coma cluster of galaxies in a 
previous attempt (Caimmi 2002), and for 
this reason our attention shall be restricted
to elliptical galaxies, where the boundaries
of the dark matter halo and the baryonic 
ellipsoid may safely be idealized as similar
and similarly placed.    

\section{Application to elliptical galaxies}
\label{apga}

According to current cosmological scenarios
(e.g., Navarro et al. 1997), density perturbations
at recombination epoch ($z\approx1400$) initially
expand with the universe, turn around, collapse,
and finally virialize (at least in their inner and
denser regions).   Just after virialization has been
attained, baryonic and (dissipationless) non baryonic 
matter are expected to fill the same volume, and
described by the same density profile.
At late times the situation changes, as energy
dissipation into heat within the gas subsystem
makes it undergo further contraction, whereas the
(non baryonic) dark matter does not appreciably
change in extension.   

Virialized density perturbations, such as elliptical
galaxies and clusters of galaxies, may safely be 
idealized as two homeoidally striated, similar and 
similarly placed, density profiles with a central 
cusp.   Our attention will be focused here on (giant)
elliptical galaxies.

\subsection{General considerations and main assumptions}
\label{geco}

A recent investigation performed on an optically
complete sample of 42 (giant) elliptical galaxies,
for which X-ray gas temperatures and central
stellar velocity dispersions were determined
(Davis \& White 1996), has shown evidence that
(giant) elliptical galaxies contain substantial
amounts of dark matter in general 
(Loewenstein \& White 1999, hereafter quoted as
LW99).   Accordingly, more than about
20\% and 39\%-85\% of the total mass
within one and six optical radii, respectively,
is in form of (non baryonic) dark matter,
depending on the stellar density profile and
observed value of X-ray gas temperature and
central stellar velocity dispersion.   The
comparison between the velocity dispersion
distributions for the dark matter and the
stars, assuming isotropic orbits, shows
that the dark matter is dynamically
``hotter'' than the stars, by a factor
1.4-2 (LW99).

The above investigation cannot be considered
as conclusive in favour of the existence of
dark matter haloes hosting (giant) elliptical
galaxies.   In fact, it has been pointed out
that the attenuation (in particular, the
scattering) by dust grains has the same effect
on the stellar kinematics as a dark matter
halo (Baes \& Dejonghe 2001).   On the other
hand, current cosmological scenarios (CDM,
$\Lambda$CDM) predict dark matter haloes hosting
elliptical galaxies, as well as spiral
galaxies, in the latter case supported by 
empirical evidence (e.g., flat rotation curves
well outside optical radii).   For this
reason, we assume that (giant) elliptical
galaxies are also embedded within dark
matter haloes.

An analysis on the evolution of the
physical properties of cosmological
baryons at low redshifts ($z\appleq
5$) has recently been performed
(Valageas et al. 2002), which (i)
yields robust model-independent
results that agree with numerical
simulations; (ii) recovers the fraction
of matter within different phases and
the spatial clustering computed by
numerical simulations; (iii) predicts
a soft X-ray background due to the
``warm'' intergalactic medium component,
that is consistent with observations.
The related baryon fraction in the 
present universe is found to be  
7\% in hot gas, 24\% in the warm
intergalactic medium, 38\% in the
cool intergalactic medium, 9\% within
star-like objects and, as a still
unobserved component, 22\% of dark
baryons associated with collapsed
structures, with a relative uncertainty
no larger than 30\% on these numbers.
Then the amount of still undetected
baryons is about one fifth of the total,
one fourth of the observed baryons
(intergalactic medium, stellar components,
and hot gas), and at least twice the
stellar-like component.

Given a typical (giant) elliptical
galaxy, a natural question is to what
extent the distribution of undetected
baryons influences the ``temperature''
of both stellar and (non baryonic) dark
matter.   Towards this aim, we make the 
following main assumptions: (a) the
stellar and the dark matter distributions
are described by homeoidally striated,
similar and similarly placed, H and NFW
density profiles, respectively; (b)
undetected baryons trace dark matter
haloes; and (c) the virial theorem in 
tensor form holds for each subsystem.
Let us discuss it briefly.

Elliptical galaxies embedded within
dark matter haloes, may safely be
idealized as two homeoidally striated,
similar and similarly placed, density
profiles with a central cusp.   According
to recent investigations (e.g., LW99),
viable representations for the outer,
non baryonic, and the inner, baryonic subsystem,
are NFW and H density profiles, respectively.
The above mentioned mass distributions were
found to be self-consistent, in a parameter
range of interest, with regard to the non
negativity of the distribution function 
(LW99) by use of a theorem due to Ciotti
\& Pellegrini (1992).

If undetected baryons in (giant) elliptical
galaxies are present as hot gas, the gaseous
subsystem is expected to be less concentrated
than the stellar one, as in the Coma cluster 
of galaxies (e.g., Briel et al. 1992).   In
this respect, the simplest assumption is that
undetected baryons trace the related, dark
matter halo, i.e. are described by NFW density 
profiles.

The typical velocity dispersion components,
deduced by use of the virial theorem in 
tensor form, are global quantities, related
to the potential-energy tensors of the
subsystem as a whole, and so, by construction,
independent of the specific orbital distribution
of the particles.   This important property,
however, is also a weakness of the virial
theorem in tensor form, when it is used to
discuss velocity dispersion components
measured in the central region of a galaxy.
In fact, it is well known that the related
values can be significantly different for
structurally identical subsystems (and so
characterized by identical virial velocity
dispersion components), due to different
orbital structures (e.g., de Zeeuw \&   
Franx 1991).   When using central
velocity dispersion components, an approach
based on Jeans equations (even though still
questionable) is to be preferred (e.g.,
Ciotti \& Lanzoni 1997; LW99).   On the 
other hand, a comparison
between the results obtained by use of either
above mentioned methods, may provide additional
support to both of them and/or useful 
indications on the nature of the problem
under investigation.

Strictly speaking, the central velocity
dispersions (along the line of sight) in
elliptical galaxies, which are deduced
from observations, should be scaled to
the corresponding typical values which 
make the virial theorem in tensor form
hold (hereafter referred to, in general,
as the virial velocity dispersions).
Both observational evidence (e.g.,
Gerhard et al. 2001) and theoretical
arguments (e.g., Nipoti et al. 2002)
point towards the existence of dynamical
homology in elliptical galaxies.   In
particular, a linear relation is found
between a local parameter, averaged 
central velocity dispersion, and a
global parameter, inferred maximum 
circular velocity, $\sigma_{0.1}=(2/3)
(v_c)_{max}$ (Gerhard et al. 2001).
Accordingly, the central velocity
dispersion components are expected
to be proportional to the virial
velocity dispersion components.
Then it is assumed that the
related proportionality factor is of the
order of unity.

In fact, typical peculiar velocity
component distributions within (giant)
elliptical galaxies show a maximum
which is rapidly attained in the central 
region (at about 1 kpc), and a slow 
decrease occurs moving outwards (no 
more than about 13\% the maximum at
about 10 kpc), at least in the case
of isotropic orbits; see e.g., LW99.
Accordingly, both the central and the
virial velocity dispersion components
are expected to be of comparable order,
slightly less than the maximum of the
peculiar velocity component distribution.
On the other hand, most elliptical galaxies 
are moderately radially anisotropic (e.g.,
Gerhard et al. 2001), and the related 
variation in central velocity dispersion
(an increase for increasing $\sigma_r^2$
and vice versa) is also expected to be
moderate.
		     
\subsection{Input parameters, specific 
assumptions, and results}
\label{inpa}

According to the above assumptions, a
(giant) elliptical galaxy is idealized
as two homeoidally striated, similar and
similarly placed matter distributions,
where the star and non baryonic subsystem
are described by H and NFW density 
profiles, respectively.   From this point
on, the inner and the outer subsystem
shall be labelled as $\ast$, $D$, instead
of $i$, $j$, respectively.   Let us suppose,
at the moment, that undetected baryons
are absent.

Following LW99, we assign to the stellar
subsystem a scaling radius $r_\ast^\dagger=
(9/20)r_{eff}$, and a truncation radius
$R_\ast=6~r_{eff}$, where $r_{eff}$ is the
optical effective radius.   Then the 
truncation, scaled radius, by use of 
Eq.\,(\ref{eq:CsiR}), is:
\begin{equation}
\label{eq:Csio}
\Xi_\ast=\frac{R_\ast}{r_\ast^\dagger}=\frac{40}3~~;
\end{equation}
independent of the optical, effective radius.
When the calculation of the scaling and the
truncation radius is needed, the typical
values:
\begin{lefteqnarray}
\label{eq:ref}
&& r_{eff}=5.04~h^{-1}{\rm kpc}~~; \\
\label{eq:ri}
&& r_\ast^\dagger=2.268~h^{-1}{\rm kpc}~~; \\
\label{eq:Ri}
&& R_\ast=30.24~h^{-1}{\rm kpc}~~;
\end{lefteqnarray}
are used, where $h$ is the dimensionless value
of the present-day Hubble parameter, normalized 
to $H_0=100$ km~s$^{-1}$~Mpc$^{-1}$.

Turning our attention to the dark matter halo,
and following again LW99, we assume for the
truncation radius and the truncation, scaled
radius, the typical values:
\begin{lefteqnarray}
\label{eq:Rj}
&& R_D=256~h^{-1}{\rm kpc}~~; \\
\label{eq:Csjo}
&& \Xi_D=10~~;
\end{lefteqnarray}
which, using Eq.\,(\ref{eq:CsiR}), yield
the following, typical value for the
scaling radius:
\begin{equation}
\label{eq:rj}
r_D^\dagger=25.6~h^{-1}{\rm kpc}~~;
\end{equation}
in addition, the dark halo appears to be  
less concentrated than the stellar ellipsoid,
i.e. $\Xi_D<\Xi_\ast$ as expected from energy
dissipation due to e.g., inelastic collisions
between pre-stellar clumps.

The mass ratio of non baryonic to baryonic
matter related to a (giant elliptical)
galaxy, equals the ratio of non baryonic 
to baryonic mean density within the volume
of the system, $\bar{\rho}_D(R_D)/\bar{\rho}
_\ast(R_D)$, or the ratio of non baryonic to
baryonic density parameter, $\Omega_D/
\Omega_b$, as (e.g., Caimmi 2002):
\begin{equation}
\label{eq:mo}
m_{D,b}=\frac{M_D}{M_b}=\frac{\bar{\rho}_D(R_D)}
{\bar{\rho}_b(R_D)}=\frac{\Omega_D}{\Omega_b}
=\frac{\Omega_M}{\Omega_b}-1~~;
\end{equation}
where $\Omega_M=\Omega_D+\Omega_b$ is the
(total) matter density parameter, and
$\Omega_b=(0.05\mp0.01)(2h)^{-2}$ fits
to a good extent data on primordial 
nucleosynthesis (e.g., White \& Fabian 1995).
Accordingly, we assume the following value
for the baryonic matter density parameter:
\begin{equation}
\label{eq:Ombo}
\Omega_b=0.0125~h^{-2}~~;
\end{equation}
for a standard $\Lambda$CDM cosmological
model with $\Omega_M=0.3$, $\Omega_\Lambda
=0.7$, the combination of Eqs.\,(\ref{eq:mo}) 
and (\ref{eq:Ombo}) yields (Caimmi 2002):
\begin{equation}
\label{eq:mho}
m_{D,b}=24~h^2-1~~;
\end{equation}
where $5\le m_{D,b}\le23$ as $0.5\le h\le1$.

The mass of each subsystem, due to Eq.\,(\ref
{eq:mo}), keeping in mind that $M_b=M_\ast$ in 
the case under discussion, is:
\begin{equation}
\label{eq:Mij}
M_\ast=\frac {M_T}{1+m_{D,\ast}}~~;\qquad M_D=\frac
{m_{D,\ast}M_T}{1+m_{D,\ast}}~~;
\end{equation}
where $M_T=M_\ast+M_D$ is the total mass, which has
to be related to the virial mass, i.e. the mass
within the virial radius of the corresponding
density perturbation.

Taking $h=2^{-1/2}$ as fiducial value, and
leaving $\Omega_M=0.3$, $\Omega_\Lambda=0.7$,
Eqs.\,(\ref{eq:ri}), (\ref{eq:Ri}), (\ref
{eq:Rj}), (\ref{eq:rj}), (\ref{eq:Ombo}), and 
(\ref{eq:mho}) yield:
\begin{leftsubeqnarray}
\slabel{eq:RrOma}
&& r_\ast^\dagger=3.21~{\rm kpc}~~;\qquad R_\ast=
42.77~{\rm kpc}~~; \\
\slabel{eq:RrOmb}
&& r_D^\dagger=36.20~{\rm kpc}~~;\qquad R_D=
362.04~{\rm kpc}~~; \\
\slabel{eq:RrOmc}
&& \Omega_b=0.025~~;\qquad m_{D,\ast}=11~~;
\label{seq:RrOm}
\end{leftsubeqnarray}
and the combination of Eqs.\,(\ref{eq:mdyd}), 
(\ref{eq:Csiy}), (\ref{eq:RrOma}), and (\ref
{eq:RrOmb}) produces:
\begin{equation}
\label{eq:yyo}
y^\dagger=11.29~~;\qquad y=8.47~~;
\end{equation}
the additional assumption that typical
dark matter haloes hosting (giant)
elliptical galaxies have masses equal 
to $5~10^{12}{\rm M}_\odot$, using 
Eqs.\,(\ref{eq:Mij}) yields:
\begin{equation}
\label{eq:Mijo}
M_\ast=4.54~10^{11}{\rm M}_\odot~~;\quad M_D=5~10^{12}
{\rm M}_\odot~~;\quad M_T=5.45~10^{12}{\rm M}_\odot~~;
\end{equation}
which completes the definition of the input
parameters.

The virial theorem in tensor form may be written
for each subsystem separately (e.g., Brosche
et al. 1983; Caimmi \& Secco 1992) as:
\begin{equation}
\label{eq:viruv}
[(E_{uv})_{vir}]_{pq}+2[(E_u)_{kin}]_{pq}=0~~;
\qquad u=\ast,D~~;\qquad v=D,\ast~~;
\end{equation}
where $[(E_u)_{kin}]_{pq}$ represents the
kinetic-energy tensor of $u$th subsystem.
Let us define a typical velocity, $v_u
\equiv[(v_u)_1, (v_u)_2, (v_u)_3]$, as:
\begin{equation}
\label{eq:Tu}
[(E_u)_{kin}]_{pq}=\frac12\delta_{pq}M_u
(v_u)_p(v_u)_q~~;\qquad u=\ast,D~~;
\end{equation}
then the combination of Eqs.\,(\ref{eq:Ma}), 
(\ref{seq:Uij}), (\ref{eq:viruv}), and (\ref
{eq:Tu}) yields:
\begin{equation}
\label{eq:vaMu}
\frac1{\epsilon_{p2}\epsilon_{p3}A_p}\frac
{(v_u)_p^2(a_u^\dagger)_1}{GM_u^\dagger}=
(\nu_u)_{mas}(\nu_{uv})_{vir}~~;\qquad u=
\ast,D~~;\qquad v=D,\ast~~;
\end{equation}
and the additional assumption of dominant
random motions implies a velocity dispersion
along the line of sight, $(\sigma_u)_p=(v_u)
_p$ in the special case where a
principal axis points towards the observer,
or using Eq.\,(\ref{eq:vaMu}):
\begin{equation}
\label{eq:saMu}
\frac{(\sigma_u)_p}{(\epsilon_{p2}\epsilon_{p3}
A_p)^{1/2}}=
\left[\frac{GM_u^\dagger}{(a_u^\dagger)_1}
(\nu_u)_{mas}(\nu_{uv})_{vir}\right]^{1/2}~~;
\qquad u=\ast,D~~;\qquad v=D,\ast~~;
\end{equation}
which has to be compared with its counterpart 
deduced from the observations.   It is apparent
that, in the absence of rotation, isotropy of 
peculiar velocity distribution occurs only for 
spherical configurations ($E0$), where $\epsilon
_{p2}\epsilon_{p3}A_p=2/3$, while the larger
extent of anisotropy is attained for the most
flattened (allowed) configurations ($E7$),
where $\epsilon_{21}=\epsilon_{31}=0.3$;
$A_1\approx0.19$; $A_2=A_3\approx0.90$;
$a_1\ge a_2\ge a_3$.

It is worth remembering that the axis ratios
and the major semiaxis, which appear in 
Eqs.\,(\ref{eq:vaMu}) and (\ref{eq:saMu}),
are related to the intrinsic configuration.
In doing the calculations, the scaling and
truncation radius will be taken as representative
of the major semiaxes of the corresponding,
isopycnic surface.
Using the results found in Subsects.\,\ref
{NFW} and \ref{H}, together with Eqs.\,(\ref
{seq:RrOm}), (\ref{eq:yyo}), (\ref{eq:Mijo}), 
and (\ref{eq:saMu}), the velocity dispersion
along the line of sight (assumed to coincide
with the direction of a principal axis) takes
the value:
\begin{equation}
\label{eq:saMij}
\frac{(\sigma_\ast)_p}{(\epsilon_{p2}\epsilon_{p3}
A_p)^{1/2}}=
287~{\rm km~s}^{-1}~~;\qquad\frac{(\sigma_D)_p}
{(\epsilon_{p2}\epsilon_{p3}A_p)^{1/2}}=225~
{\rm km~s}^{-1}~~;
\end{equation}
with regard to the stellar subsystem, it
is consistent with a typical value $\sigma
_\ast=250$~km~s$^{-1}$ (e.g., LW99; Treu et al.
2001) except in the limiting case of 
(intrinsic) $E7$ configurations with the 
major axis pointing towards the observer.

The above results hold provided the baryonic
subsystem is mainly in form of stars.   Let
us take into consideration a different 
scenario, where a less concentrated gaseous subsystem
than the stellar one is also present,
as in the Coma cluster of galaxies 
(e.g., Briel et al. 1992), and assume the same
mass distribution as in the non baryonic
matter (Caimmi 2002).   Accordingly, a
(giant) elliptical galaxy may safely be
conceived as formed by an inner subsystem 
made of stars and an outer subsystem made
of gas and non baryonic matter.   As the
amount of baryonic and non baryonic
matter have to remain unchanged, the
inner and the outer subsystem are less
and more massive, respectively, than
in absence of undetected baryons.   Again,
it is assumed that the related mass
distributions correspond to a H and NFW
density profile, respectively.     Then
we define the mass ratio:
\begin{equation}
\label{eq:msf}
m_{g,\ast}=\frac{M_g}{M_\ast}~~;
\end{equation}
where $M_\ast$ and $M_g$ are the mass
of the stellar and gaseous subsystem,
respectively.    Bearing in mind that 
$M_\ast+M_g=M_b$, where $M_b$ is the total 
mass in baryons,
the following relations are easily derived:
\begin{leftsubeqnarray}
\slabel{eq:Mijsa}
&& M_\ast=\frac{M_b}{1+m_{g,\ast}}~~;\quad M_D+M_g
=\left(1+\frac{m_{g,\ast}}{m_{D,\ast}}\right)M_D
~~; \\ 
\slabel{eq:Mijsb}
&& m_{D,\ast}+m_{g,
\ast}=(1+m_{g,\ast})m_{D,b}+m_{g,\ast}~~;
\label{seq:Mijs}
\end{leftsubeqnarray}
and the repetition of the above procedure
with $M_\ast=M_b$, $M_D$, $m_{D,\ast}=m_
{D,b}$, replaced by $M_\ast=M_b-M_g$,
$M_D+M_g$, $m_{D,\ast}+m_{g,\ast}$,
respectively, allows the specification 
of the input parameters.

Some output parameters, such as profile
factors involving both the inner and the
outer matter distribution, and velocity dispersions
along the line of sight, depend on the mass
of each subsystem, and then on the mass 
ratio of gaseous to stellar subsystem,
$m_{g,\ast}$, via Eqs.\,(\ref{seq:Mijs}).   
The following results are found:
\begin{lefteqnarray}
\label{eq:sin}
&& \frac{(\sigma_\ast)_p/({\rm km~s}^{-1})}
{(\epsilon_{p2}\epsilon_{p3}A_p)^{1/2}}=
246, 223, 208, 198, 190, 184~~; \\
\label{eq:sjn}
&& \frac{(\sigma_D)_p/({\rm km~s}^{-1})}
{(\epsilon_{p2}\epsilon_{p3}A_p)^{1/2}}=
217, 213, 211, 210, 208, 208~~; \\
\label{eq:fn}
&& m_{g,\ast}=0.5,1.0,1.5,2.0,2.5,3.0~~;
\end{lefteqnarray}
where $m_{g,\ast}$ has been considered as 
a free parameter, in the range of interest.

It is apparent that the outer subsystem
(gas plus non baryonic matter) is 
dynamically ``hotter''
than the inner, provided the gas mass
fraction exceeds the star mass fraction
by a factor of about 3/2.   The
velocity dispersion along the line of
sight decreases as $m_{g,\ast}$ increases, and
a similar trend occurs, via Eqs.\,(\ref
{eq:saMu}), (\ref{eq:msf}), and (\ref
{seq:Mijs}), as the total mass decreases.

On the other hand, the comparison (under
the assumption of isotropic orbits) between
the peculiar velocity component distributions
related to the stellar ellipsoid and the dark
matter halo, shows that the latter is 
dynamically ``hotter'' than the former (LW99).
The reasons for this discrepancy may be due
to (i) the different method used, with respect to 
the virial theorem in tensor form, in 
determining the ``temperature'' of the
stellar ellipsoid and the dark matter halo
(unfortunately, the derivation of the peculiar
velocity component distribution is not outlined
in LW99), and (ii) the different values of the
mass ratio between the outer and inner
subsystem in absence ($m_{D,\ast}=11$) and in presence
($m_{D,\ast}+m_{g,\ast}=29)$, according to Eqs.\,(\ref
{seq:Mijs}), near equally ``hot'' stellar
ellipsoid and dark matter halo) of undetected
baryons: values $m_{D,\ast}+m_{g,\ast}>29$ would 
allow dynamically
``hotter'' dark matter haloes, apart from the
nature of their constituents.

According to a recent investigation
(Valageas et al. 2002), the ratio of
undetected baryons associated with
collapsed structures to star-like
objects, in the present universe,
attains a value which
is close to 2.   If undetected
baryons and stars are present to 
a similar extent within (giant)
elliptical galaxies, then $m_{g,\ast}\approx
2$ and dark haloes, according to
Eqs.\,(\ref{eq:sin}), (\ref{eq:sjn}), 
and (\ref{eq:fn}), are dynamically
``hotter'' than
stellar ellipsoids, in the case 
under discussion.   
In this view, both the observation that the temperature of the
extended hot gas exceeds the central stellar temperature, and
the fact that the non baryonic matter is dynamically ``hotter''
than the stars (e.g., LW99), are a reflection of the presence
of undetected baryons, which trace the dark halo and are about 
twice as massive as the stellar ellipsoid.

Though (giant)
elliptical galaxies host a surprisingly
large amount of interstellar dust (up
to a several ten million solar masses),
most of it believed to be distributed
diffusely over the galaxy (e.g., Baes
\& Dejonge 2001), this is still a
negligible fraction of the total mass
of the baryonic subsystem, such that
a substantial amount of interstellar
baryons within 
(giant) ellipticals, if it really 
exists, still has to be detected.

\section{Concluding remarks}\label{core}

A general theory of homeoidally striated
ellipsoids (e.g., Roberts 1962; Chandrasekhar
1969; C93), where no divergence occurs in the
density profile, has been adapted to cuspy
density profiles.   An
explicit calculation of the related, physical 
parameters implies the specification of
the density profile, which is equivalent to
the knowledge of: (i) the functional dependence
of a scaled density, $f=\rho/\rho^\dagger$, on a
scaled radius, $\xi=r/r^\dagger$; (ii) a boundary 
condition, i.e. $f(1)=1$; (iii) two independent
parameters, i.e. a scaling density, $\rho^\dagger$,
and a scaling radius, $r^\dagger$; (iv) a truncated,
scaled radius, $\Xi$.    The latter requirement
is due to the fact, that the systems under 
consideration exhibit a null density at an
infinite radius where, on the other hand, the 
total mass may attain a divergent value.    In
addition, an infinity of density profiles in
the physical space, $({\sf O}r\rho)$, is
represented by a single density profile in
the abstract space, $({\sf O}\xi f)$, for
any selected choice of exponents, $(\alpha,
\beta,\gamma)$, appearing in Eq.\,(\ref
{eq:runi}).
 
Potential-energy tensors involving both one
and two, homeoidally striated density profiles,
where the boundaries are similar and similarly
placed, have been expressed in terms of integrals
on the mass distribution.   Explicit calculations
have been performed for both NFW and MOA density
profiles, which satisfactorily fit the results
of high-resolution simulations for dark matter
haloes (e.g., Fukushige
\& Makino 2001; Klypin et al. 2001), and for H
density profiles, which closely approximate the
de Vaucouleurs $r^{1/4}$ law for elliptical
galaxies (e.g., Hernquist 1990; Holley-Bockelmann
et al. 2001).

The virial theorem in tensor form, related to a 
two-component system, has been 
expressed for each subsystem, and applied
to giant elliptical galaxies.   The predicted velocity
dispersion along the line of sight, in
the limiting case where a principal axis points
towards the observer, has been
found to be consistent with the data except
for (intrinsic) $E7$ configurations, when the 
major axis points towards the observer.   

The suggestion that dark matter haloes host an 
amount of undetected baryons as massive as 
about twice the stellar subsystem (Valageas 
et al. 2002), together with the assumption 
that undetected baryons trace non baryonic 
matter therein, has produced two main 
consequences, namely (i) predicted velocity 
dispersions along the line of sight are lower 
than in absence of undetected baryons, and 
(ii) dark matter haloes are dynamically 
``hotter'' than stellar 
ellipsoids, the transition occurring when 
the amount of undetected baryons is about 
one and a half times the stellar subsystem.

In this view, both the observation that the temperature of the
extended hot gas exceeds the central stellar temperature, and
the fact that the non baryonic matter is dynamically ``hotter''
than the stars (e.g., LW99), are a reflection of the presence
of undetected baryons, which trace the dark halo and are about 
twice as massive as the stellar ellipsoid.

\begin{acknowledgements}
We are indebted to an anonymous referee for critical
comments which made substantial improvements to an
earlier version of the manuscript.   The
analytical integrations needed in the current paper
were helped substantially by visiting the internet site:
``HTTP://INTEGRALS.WOLFRAM.COM/INDEX.CGI''.   This
is why we are deeply grateful to the Wolfram staff,
in particular to Eric Weisstein, and wish to 
acknowledge all the facilities encountered 
therein. 
\end{acknowledgements}

\appendix
\section{A procedure for fitting simulated
density profiles
}\label{espo}

Simulated, spherically averaged, virialized
dark matter haloes, allow the knowledge of the
density profile, the virialized mass, and the 
virialized
radius, where the last two may safely be taken
as representative of the whole mass and the
whole radius, respectively (e.g., Cole \&
Lacey 1996; Navarro et al. 1997; Fukushige
\& Makino 2001; Klypin et al. 2001).   A plot of
the density profile on the logarithmic plane,
$({\sf O}\log\xi\log f)$, necessarily implies
use of dimensionless coordinates, defined by
Eqs.\,(\ref{eq:rhoa}) and (\ref{eq:csir}),
namely:
\begin{equation}
\label{eq:coa}
f(\xi)=\frac\rho{\rho^\dagger}~~;\quad\xi=\frac r{r^\dagger}~~;
\end{equation}
where the choice of the scaling density, $\rho^\dagger$, 
and the scaling radius, $r^\dagger$, is, in principle,
arbitrary.   A different choice of the above
mentioned parameters, say from $(\rho^\dagger,r^\dagger)$ to
$[(\rho^\prime)^\dagger,(r^\prime)^\dagger]$, makes simulated
points shift by a factor, $\log[(\rho^\prime)^\dagger/
\rho^\dagger]$, along the $\log f$ axis, and by a factor, 
$\log[(r^\prime)^\dagger/r^\dagger]$, along the $\log\xi$ axis.

In fitting simulated density profiles, our
attention shall be restricted to the family 
of functions (e.g., Hernquist 1990; Zhao 1996):
\begin{equation}
\label{eq:Zad}
f(\xi)=\frac A{\xi^\gamma(1+\xi^\alpha)^\chi}~~;
\quad\chi=\frac{\beta-\gamma}\alpha~~;
\end{equation}
which has been deduced from Eqs.\,(\ref{eq:runi}) 
and (\ref{eq:coa}), provided the boundary 
condition:
\begin{equation}
\label{eq:bc}
\rho(r^\dagger)=\rho^\dagger f(1)=\rho^\dagger~~;\quad f(1)=1~~;
\quad\rho^\dagger=\rho(1)~~;
\end{equation}
holds, according to Eq.\,(\ref{eq:rhoa}).   The
combination of Eqs.\,(\ref{eq:Zad}) and
(\ref{eq:bc}) yields:
\begin{equation}
\label{eq:A}
f(1)=A~2^{-\chi}=1~~;\quad A=2^\chi~~;
\end{equation}
and Eq.\,(\ref{eq:Zad}) takes the final
form:
\begin{equation}
\label{eq:Zbc}
f(\xi)=\frac{2^\chi}{\xi^\gamma(1+\xi^\alpha)
^\chi}~~;
\end{equation}
which is equivalent to:
\begin{equation}
\label{eq:lgZ}
\log f=\chi\log2-\gamma\log\xi-\chi\log(1+\xi
^\alpha)~~;
\end{equation}
due to Eq.\,(\ref{eq:bc}), the three-parameter
curve represented by Eq.\,(\ref{eq:lgZ}) remains
unchanged in the logarithmic plane $({\sf O}\log
\xi\log f)$, for different choices of the scaling
parameters, $(\rho^\dagger,r^\dagger)$.   On the other hand,
due to Eq.\,(\ref{eq:coa}), any point $(\xi,f)$
on the curve represents different points $(\xi r^
\dagger,f\rho^\dagger)$ on (one or more) physical density 
profiles, for different choices of the scaling
parameters, $(\rho^\dagger,r^\dagger)$.

In the limit of negligible values of the 
independent variable, $\xi$, with respect to
unity, Eq.\,(\ref{eq:lgZ}) reduces to:
\begin{equation}
\label{eq:fcmn}
\log f=\chi\log2-\gamma\log\xi~~;\quad\xi\ll1~~;
\end{equation}
which represents, in the logarithmic plane, a
straight line with slope equal to $-\gamma$ 
and intercept equal to $\chi\log2$.

In the limit of preponderant values of the 
independent variable, $\xi$, with respect to
unity, Eq.\,(\ref{eq:lgZ}) reduces to:
\begin{equation}
\label{eq:fcmx}
\log f=\chi\log2-\beta\log\xi~~;
\quad\xi\gg1~~;
\end{equation}
which represents, in the logarithmic plane, a
straight line with slope equal to $-\beta=-(\gamma+
\chi\alpha)$ and intercept equal to $\chi\log2$.

\begin{figure}
   \centering
   \resizebox{\hsize}{!}{\includegraphics{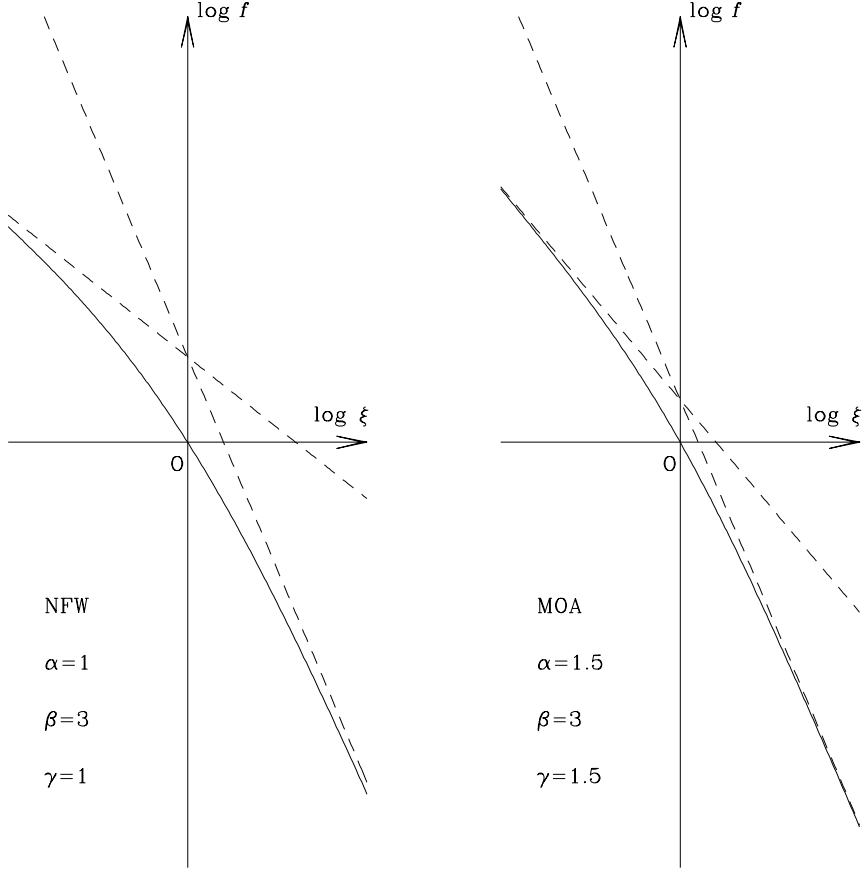}} 
\caption{Representation, in the logarithmic plane, of 
NFW (left) and MOA (right) density profiles (full 
curves), with their asymptotes (dashed lines).   
In both cases, $-1<\log\xi<+1$ and $-3<\log f<+3$.}
\label{f:lMN}
    \end{figure}

The straight lines under discussion have
coinciding intercepts i.e. they meet on the
vertical axis and, in addition, represent
the asymptotes of the curve, expressed by
Eq.\,(\ref{eq:lgZ}).   The special cases
related to NFW and MOA density profiles,
are plotted in Fig.\,\ref{f:lMN}.

The above results hold for $\alpha>0$.    The case
$\alpha<0$ makes the asymptotes change one into the
other.   The limiting case $\alpha=0$ makes the
asymptotes coincide i.e. the curve reduces to a 
straight line.


A change of variables:
\begin{equation}
\label{eq:cvar}
x=\log\xi~~;\qquad y=\log f~~;
\end{equation}
translates Eq.\,(\ref{eq:lgZ}) into the equivalent form:
\begin{equation}
\label{eq:xy0}
y=\chi\log2-\gamma x-\chi\log[1+\exp_{10}(\alpha x)]~~;
\end{equation}
where, in general, $\exp_u(x)=u^x$, and $\exp
(x)={\rm e}^x$, according to the standard notation.   
The derivatives up to the third order are:
\begin{leftsubeqnarray}
\slabel{eq:xy1}
&& \frac{\diff y}{\diff x}=-\beta+\frac{\chi\alpha}
{1+\exp_{10}(\alpha x)}~~; \\
\slabel{eq:xy2}
&& \frac{\diff^2y}{\diff x^2}=-\chi\alpha^2\ln(10)\frac
{\exp_{10}(\alpha x)}{[1+\exp_{10}(\alpha x)]^2}~~; \\
\slabel{eq:xy3}
&& \frac{\diff^3y}{\diff x^3}=-\chi\alpha^3\ln^2(10)\exp_
{10}(\alpha x)\frac{1-\exp_{10}(\alpha x)}{[1+\exp_{10}
(\alpha x)]^3}~~;
\label{seq:xy}
\end{leftsubeqnarray}
and the particularization of Eq.\,(\ref{eq:xy1}) 
to $x=0$ i.e. $\xi=1$ or $r=r^\dagger$, reads:
\begin{equation}
\label{eq:xy00}
\left(\frac{\diff y}{\diff x}\right)_{x=0}=-\frac12
(\gamma+\beta)~~;
\end{equation}
in the special case of NFW density profile, the 
right-hand side member reduces to -2, as pointed 
out by Bullock et al. (2001).    On the other side,
Eq.\,(\ref{eq:xy3}) shows that the third derivative 
equals zero at the same point, $x=0$ i.e. $\xi=1$ 
or $r=r^\dagger$.    Accordingly, density profiles
expressed by Eqs.\,(\ref{eq:coa}) and (\ref{eq:Zad}),
are characterized by a maximum variation in slope,
in the logarithmic plane $({\sf O}~\log\xi~\log f)$, 
at that
point where $x=0$ i.e. $\xi=1$ or $r=r^\dagger$, and
the slope is expressed by Eq.\,(\ref{eq:xy00}).

Bearing in mind that fixed points on a physical
density profile change their position on the
scaled density profile, for different choices of 
the scaling parameters, $(\rho^\dagger,r^\dagger)$, according
to Eq.\,(\ref{eq:coa}), a possible procedure in
fitting density profiles of simulated, virialized, 
dark matter haloes, could be the following.
\begin{description}
\item[\rm{(i)}] Assume that self-similar,
universal density profiles, expressed by Eq.\,(\ref
{eq:Zbc}), are representative of virialized, dark matter
haloes (e.g., Cole \& Lacey 1996; Navarro et al. 1997; 
Fukushige \& Makino 2001; Klypin et al. 2001).
\item[\rm{(ii)}] Select two special choices of scaling 
parameters, $[(\rho^\dagger)^-,(r^\dagger)^-]$ and 
$[(\rho^\dagger)^+,(r^\dagger)^+]$,
which make simulated points well shifted on the left and
on the right, respectively, i.e. (expected to be) 
sufficiently close to either asymptote.
\item[\rm{(iii)}] Select the best pair of deduced
asymptotes, which meet together on the vertical axis.
\item[\rm{(iv)}] Calculate the slope and 
the intercept of either asymptote.
\item[\rm{(v)}] Calculate the exponents, $(\alpha, \beta,
\gamma)$, via Eqs.\,(\ref{eq:Zad}), (\ref{eq:fcmn}) and
(\ref{eq:fcmx}).
\end{description}

In substance, the above procedure relies on the 
possibility, that a least-square fit related to
a curve, represented by Eq.\,(\ref{eq:lgZ}), may
be reduced to a least-square fit related to
a straight line, represented by either Eq.\,(\ref
{eq:fcmn}) or Eq.\,(\ref{eq:fcmx}).

We hope that, in fitting self-similar, universal 
density profiles of simulated, virialized, dark matter 
haloes, with mass distribution expressed by Eq.\,(\ref
{eq:coa}), the procedure will be explained in detail.

\section{Connection between NFW and MOA density  
profiles in fitting simulated haloes}\label{proFM}

A generic density profile, belonging to the family
expressed by Eq.\,(\ref{eq:runi}), is defined by 
three exponents,
$(\alpha, \beta, \gamma)$, and is characterized by
two independent parameters e.g., a scaling density,
$\rho^\dagger$, and a scaling radius, $r^\dagger$, or any other
pair of independent parameters.   In particular, 
the total mass within the virial radius, $M=M_{200}$,
and a dimensionless parameter, $\delta$, may be
used, and their expression as a function of $\rho^\dagger$
and $r^\dagger$ are derived by fitting numerical simulations
(Fukushige \& Makino 2001, hereafter quoted as FM01).     

Then the problem is
how to compare simulated density profiles with
different, theoretical ones of the kind under 
consideration.   In doing this, our attention
shall be restricted to (a) high-resolution
simulations from FM01,
and (b) NFW and MOA theoretical, density 
profiles, but the method is quite general.

With regard to NFW density profiles, the 
best fit to numerical simulations prescribed 
by FM01:
\begin{equation}
\label{eq:fNFM}
f(\xi)=\frac{10}{(\xi/2)[1+(\xi/2)]^2}~~;
\end{equation}
is inconsistent with the curve plotted in
their Fig.\,18 (hereafter quoted as Fig.\,FM18), 
probably due to some printing
errors.   In fact, the comparison with their 
Fig.\,17 (hereafter quoted as Fig.\,FM17), 
which shows the counterpart of 
Eq.\,(\ref{eq:fNFM}) related to MOA density
profile:
\begin{equation}
\label{eq:fMFM}
f(\xi)=\frac1{\xi^{3/2}(1+\xi^{3/2})}~~;
\end{equation}
allows the following conclusions: (i) 
simulated density profiles remain unchanged,
when both the above mentioned plots have
same origin and same scale; and (ii) both
NFW and MOA density profiles, represented
therein, attain the same value as $\xi
\rightarrow+\infty$.

With these constraints, Eq.\,(\ref{eq:fNFM})
takes the form:
\begin{equation}
\label{eq:fNFMR}
f(\xi)=\frac{k_1^3}{(k_1\xi)[1+(k_1\xi)]^2}~~;
\end{equation}
where $k_1$ is a parameter which can be 
calculated from the knowledge of the 
coordinates of an arbitrary selected point
on the NFW density profile, plotted in
Fig.\,FM18.  

Following FM01, NWF and MOA density
profiles are normalized as in Eqs.\,(\ref{eq:fNFM}) 
and (\ref{eq:fMFM}), respectively, and plotted in
Figs.\,FM17 and FM18, respectively.   On the 
other hand, Eq.\,(\ref{eq:fNFM}) appears to be
inconsistent with the curve plotted in Fig.\,FM18.
To get a deeper insight into this
problem, let us start with the obvious relations:
\begin{lefteqnarray}
\label{eq:rho**}
&& \frac{\rho}{(\rho^\prime)^\dagger_{NFW}}=\frac
{\rho}{(\rho^\prime)^\dagger_{MOA}}\frac{(\rho^\prime)^\dagger_
{MOA}}{(\rho^\prime)^\dagger_{NFW}}=\rho_{**}\frac
{(\rho^\prime)^\dagger_{MOA}}{(\rho^\prime)^\dagger_{NFW}}~~;
\quad\rho_{**}=\frac\rho{(\rho^\prime)^\dagger_{MOA}}~~; \\
\label{eq:r**}
&& \frac r{(r^\prime)^\dagger_{NFW}}=\frac r{(r^\prime)^\dagger_
{MOA}}\frac{(r^\prime)^\dagger_{MOA}}{(r^\prime)^\dagger_{NFW}}=
r_{**}\frac{(r^\prime)^\dagger_{MOA}}{(r^\prime)^\dagger_{NFW}}~~;
\quad r_{**}=\frac r{(r^\prime)^\dagger_{MOA}}~~;
\end{lefteqnarray}
where $\rho_{**}$, $r_{**}$, are defined as in
FM01, and $(\rho^\prime)^\dagger$, $(r^\prime)^\dagger$, have the 
same values as prescribed therein.

Bearing in mind that any point, ${\sf P}(r_{**},
\rho_{**})$, maintains its coordinates passing from
Fig.\,FM17 to Fig.\,FM18 and vice versa, the NFW 
density profile plotted in Fig.\,FM18, reads:
\begin{leftsubeqnarray}
\slabel{eq:NFW**a}
&& \rho_{**}=\frac{k_1k_2}{(k_1r_{**})[1+(k_1r_
{**})]^2}~~; \\
\slabel{eq:NFW**b}
&& k_1=\frac{(r^\prime)^\dagger_{MOA}}{(r^\prime)^\dagger_{NFW}}~~;
\quad k_2=\frac1{k_1}\frac{(\rho^\prime)^\dagger_{NFW}}
{(\rho^\prime)^\dagger_{MOA}}~~;
\label{seq:NFW**b}
\end{leftsubeqnarray}
further inspection of Figs.\,FM17 and FM18
discloses the validity of the condition:
\begin{equation}
\label{eq:lim**}
\lim_{r_{**}\to+\infty}\frac{r_{**}^{-3/2}(1+
r_{**}^{3/2})^{-1}}{k_2r_{**}^{-1}(1+k_1r_{**})^{-2}}=1~~;
\end{equation}
which, performing the related calculations, is
equivalent to:
\begin{equation}
\label{eq:k2k1}
k_2=k_1^2~~;
\end{equation}
and the combination of Eqs.\,(\ref{eq:NFW**a}) 
and (\ref{eq:k2k1}) yields Eq.\,(\ref{eq:fNFMR}).
The comparison of Eqs.\,(\ref{eq:fNFM}), (\ref
{eq:fMFM}), with (\ref{eq:rhoN}), (\ref{eq:rhoM}),
respectively, yields:
\begin{leftsubeqnarray}
\slabel{eq:norFMa}
&& 2(\rho^\dagger)_{MOA}=(\rho^\prime)^\dagger_{MOA}~~;\quad
(r^\dagger)_{MOA}=(r^\prime)^\dagger_{MOA}~~; \\
\slabel{eq:nor**b}
&& 4(\rho^\dagger)_{NFW}=k_1^3(\rho^\prime)^\dagger_{MOA}~~;\quad
(r^\dagger)_{NFW}=k_1^{-1}(r^\prime)^\dagger_{MOA}~~;
\label{seq:nor**}
\end{leftsubeqnarray}
which translate the prescriptions from FM01,
their Eqs.\,(13) and (14), into the following:
\begin{leftsubeqnarray}
\slabel{eq:rho0a}
&& \rho^\dagger=C_\rho\delta\left(\frac M{{\rm M}_{10}}
\right)^{-1}\frac{10^{10}{\rm M}_\odot}{{\rm kpc}^3}~~; \\
\slabel{eq:rho0b}
&& (C_\rho)_{NFW}=\frac{7k_1^3}{40}~~;\quad
(C_\rho)_{MOA}=\frac7{20}~~;
\label{seq:rho0}
\end{leftsubeqnarray}
with regard to the scaling density, $\rho^\dagger$, and:
\begin{leftsubeqnarray}
\slabel{eq:r0a}
&& r^\dagger=C_r\delta^{-1/3}\left(\frac M{{\rm M}_{10}}
\right)^{2/3}{\rm kpc}~~; \\
\slabel{eq:r0b}
&& (C_r)_{NFW}=2~10^{-2/3}k_1^{-1}~~;\quad
(C_r)_{MOA}=2~10^{-2/3}~~;
\label{seq:r0}
\end{leftsubeqnarray}
with regard to the scaling radius, $r^\dagger$,
where ${\rm M}_{10}=10^{10}{\rm M}_\odot$ and 
the values of $(\rho^\dagger)_{NFW}$, $(\rho^
\dagger)_{MOA}$, 
are one fourth, one half, respectively, the 
value derived in FM01, due to the different 
normalization of the NFW, MOA, density profiles, 
adopted in the current paper.   The value of the 
dimensionless
parameter, $\delta$, is considered to reflect
an amplitude of the density fluctuation at
turnaround and, for this reason, it can be
thought of as constant during the evolution
of a halo (e.g., Cole \& Lacey 1996; Navarro
et al. 1997; FM01).   From the standpoint of 
top-hat, spherical density perturbation, it 
is related to both the mass and the peak
height.

At this stage, what still remains to be done is the
specification of the NFW density profile 
related to the fit of simulated, dark
matter haloes from FM01 plotted in Fig.\,FM18.   
Towards this aim, let ${\sf\hat{P}}(\hat{r}_{**},
\hat{\rho}_{**})$ be a selected point on the 
curve plotted in Fig.\,FM18.
Accordingly, Eq.\,(\ref{eq:fNFMR}) reads:
\begin{equation}
\label{eq:rho^**}
\hat{\rho}_{**}=\frac{k_1^2}{\hat{r}_{**}[1+
(k_1\hat{r}_{**})]^2}~~;
\end{equation}
and the related, second-degree equation in 
$k_1$ has the solutions:
\begin{equation}
\label{eq:k1+-}
k_1^\mp=\hat{r}_{**}^{-1}\{[1\pm(\hat{\rho}_{**}
\hat{r}_{**}^3)^{1/2}]^{-1}-1\}~~;
\end{equation}
where the negative solution has no physical
meaning.   Then the acceptable solution to
Eq.\,(\ref{eq:rho^**}) is:
\begin{equation}
\label{eq:k1+}
k_1=\hat{r}_{**}^{-1}\{[1-(\hat{\rho}_{**}
\hat{r}_{**}^3)^{1/2}]^{-1}-1\}~~;
\end{equation}
where it is intended that the square root is
non negative.

The results related to a pair of selected points
on the curve plotted in Fig.\,FM18, are
listed in Table~\ref{t:FM}.    The counterpart
\begin{table}
\begin{tabular}{ccccc}
\hline
\hline
\multicolumn{1}{c}{$\hat{\rho}_{**}$} &
\multicolumn{1}{c}{$\hat{r}_{**}$}  &
\multicolumn{1}{c}{$k_1$}  &
\multicolumn{1}{c}{$k_1^2$} &
\multicolumn{1}{c}{$k_1^3$} \\
\hline
1000 & 0.005 & 2.26 & 5.11 & 11.56 \\
500 & 0.01 & 2.29 & 5.23 & 11.91 \\
\hline\hline
\end{tabular}
\caption{Specification via the parameter, $k_1$,
of the NFW density profile related to the fit of
simulated, dark matter haloes from FM01, plotted
in Fig.\,FM18.   The coordinates of the two
points, ${\sf\hat{P}}(\hat{r}_{**},\hat{\rho}_
{**})$, have been deduced from Fig.\,FM18.}
\label{t:FM}
\end{table}
of Eq.\,(\ref{eq:fNFMR}) in FM01, probably due 
to some printing 
errors, is inconsistent with the results listed
in Table~\ref{t:FM}, as it would imply
$k_1^3=10$, $k_1=0.5$, according to 
Eq.\,(\ref{eq:fNFM}).   Consistent,
even if non acceptable, choices would
be either $k_1^3=8$, $k_1=2$, or $k_1^3
=10$, $k_1=10^{1/3}=2.1544$.

\end{document}